\documentclass[letterpaper, journal]{IEEEtran}
\newif\ifshowcomment
\newif\ifnumberrevision
\newif\ifcolorrevision
\newif\ifstrikeremovel

\usepackage{textcomp}
\usepackage{url}
\usepackage{verbatim}

\usepackage{graphicx}
\usepackage{amsmath,amssymb,enumerate}
\usepackage{amsfonts}
\usepackage{mathtools}
\usepackage[mathscr]{euscript}
\usepackage{mathrsfs}
\usepackage{times}
\usepackage[usenames,dvipsnames,svgnames,table, xcdraw]{xcolor} %http://ctan.org/pkg/xcolor
\usepackage[caption=false,font=normalsize,labelfont=sf,textfont=sf]{subfig}
\usepackage{booktabs}
\usepackage{multirow}
\usepackage[ruled,vlined,linesnumbered]{algorithm2e}
\SetKw{Continue}{continue}
\SetKw{Break}{break}

\usepackage{enumitem}
\usepackage{cite} % to group references
\usepackage[utf8]{inputenc}
\usepackage{balance}
\usepackage[normalem]{ulem}
\usepackage{pifont}

\newcommand{\Zset}{\mathbb{Z}_{\geq 0}}

\newcommand{\highlightontology}[1]{\textit{#1}}
\newcommand{\tableref}{TABLE}

\title{\LARGE \bf Digital Twin-based Smart Manufacturing: Dynamic Line Reconfiguration for Disturbance Handling}

\author{Bo Fu, Mingjie Bi, Shota Umeda, Takahiro Nakano, Youichi Nonaka, \\ Quan Zhou, Takaharu Matsui, Dawn M. Tilbury, Kira Barton%

\thanks{This work was supported by Hitachi Ltd.
The work was completed before Bo Fu joined Amazon.com Services LLC.
\textit{(Corresponding author: Bo Fu.)} }

\thanks{Bo Fu was with the Department of Robotics, University of Michigan, Ann Arbor, MI 48109 USA. He is now with Amazon Robotics, North Reading, MA 01864 USA (e-mail: bofu@umich.edu).}
\thanks{Mingjie Bi was with the Department of Robotics, University of Michigan, Ann Arbor, MI 48109 USA. He is now with the Beijing Institute for General Artificial Intelligence, Beijing 100080, China (e-mail: mingjieb@umich.edu).}
\thanks{Shota Umeda, Takahiro Nakano, and Youichi Nonaka are with Hitachi, Ltd., Tokyo 100-8280, Japan (e-mail: shota.umeda.zf@hitachi.com; takahiro.nakano.tz@hitachi.com; youichi.nonaka.ym@hitachi.com).}
\thanks{Quan Zhou and Takaharu Matsui are with Hitachi America, Ltd., Farmington Hills, MI 48355 USA (e-mail: quan.zhou@hal.hitachi.com; takaharu.matsui@hal.hitachi.com).}
\thanks{Dawn M. Tilbury is with the Department of Robotics, University of Michigan, Ann Arbor, MI 48109 USA (e-mail: tilbury@umich.edu).}
\thanks{Kira Barton is with the Department of Robotics and the Department of Mechanical Engineering, University of Michigan, Ann Arbor, MI 48109 USA (e-mail: bartonkl@umich.edu).}
}

\begin{document}

% Override seps
% \setlength{\textfloatsep}{\baselineskip}
% \setlength{\floatsep}{\baselineskip}
% \setlength{\intextsep}{\baselineskip}
% \setlength{\dbltextfloatsep}{\baselineskip}
% \setlength{\dblfloatsep}{\baselineskip}

% \floatsep 1\baselineskip plus  0.2\baselineskip minus  0.2\baselineskip
% \intextsep 1\baselineskip plus 0.2\baselineskip minus  0.2\baselineskip

\maketitle
\begin{abstract}
The increasing complexity of modern manufacturing, coupled with demand fluctuation, supply chain uncertainties, and product customization, underscores the need for manufacturing systems that can flexibly update their configurations and swiftly adapt to disturbances. However, current research falls short in providing a holistic reconfigurable manufacturing framework that seamlessly monitors system disturbances, optimizes alternative line configurations based on machine capabilities, and automates simulation evaluation for swift adaptations. This paper presents a dynamic manufacturing line reconfiguration framework to handle disturbances that result in operation time changes.
The framework incorporates a system process digital twin for monitoring disturbances and triggering reconfigurations, a capability-based ontology model capturing available agent and resource options, a configuration optimizer generating optimal line configurations, and a simulation generation program initializing simulation setups and evaluating line configurations at approximately 400x real-time speed.
A case study of a battery production line has been conducted to evaluate the proposed framework. In two implemented disturbance scenarios, the framework successfully recovers system throughput with limited resources, preventing the 26\% and 63\% throughput drops that would have occurred without a reconfiguration plan. The reconfiguration optimizer efficiently finds optimal solutions, taking an average of 0.03 seconds to find a reconfiguration plan for a manufacturing line with 51 operations and 40 available agents across 8 agent types.
\end{abstract}

\def\abstractname{Note to Practitioners}
\begin{abstract}
This research is motivated by the need for an advanced framework that automates the reconfiguration of manufacturing lines in response to increasing disturbances within dynamic production environments.
Existing approaches lack a holistic framework that automates disturbance monitoring, data management, line structure optimization, and pre-implementation simulation verification.
This paper integrates all these components and develops an innovative configuration optimizer and simulator to improve the optimality and efficiency of disturbance handling.
We demonstrate the applicability of our framework through a case study in a battery production line, focusing on agent delay issues.
The approach is adaptable to a variety of disturbance scenarios in other types of manufacturing lines that share common elements such as workers, robots, machines, and auxiliary tools.
Future investigations will expand on reconfiguration considerations, including spatial constraints, agent experience levels, and tool change requirements.
\end{abstract}


\section{Introduction}\label{sec:introduction}
\begin{figure*}[t]
    \centering
	\subfloat[\label{fig:line_disturbance_handle_baseline}]{
    	\includegraphics[width=0.95\linewidth, trim=0 0 0 0, clip]{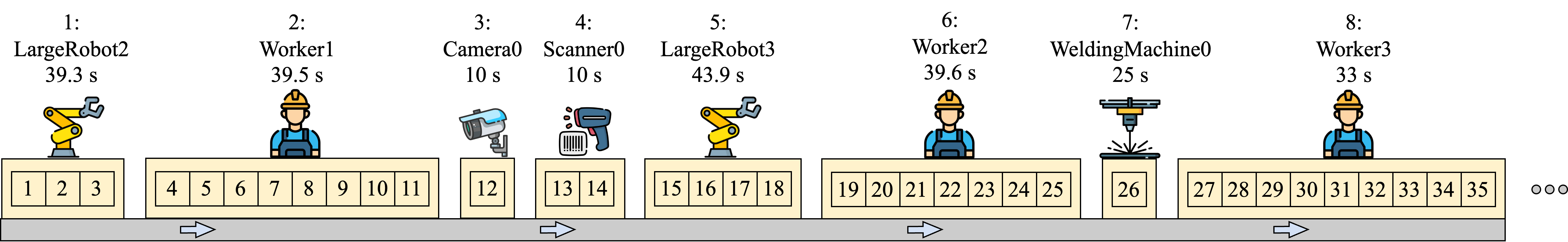}}
    \hfill
	\subfloat[\label{fig:line_disturbance_handle_adjust}]{
    	\includegraphics[width=0.95\linewidth, trim=0 0 0 0, clip]{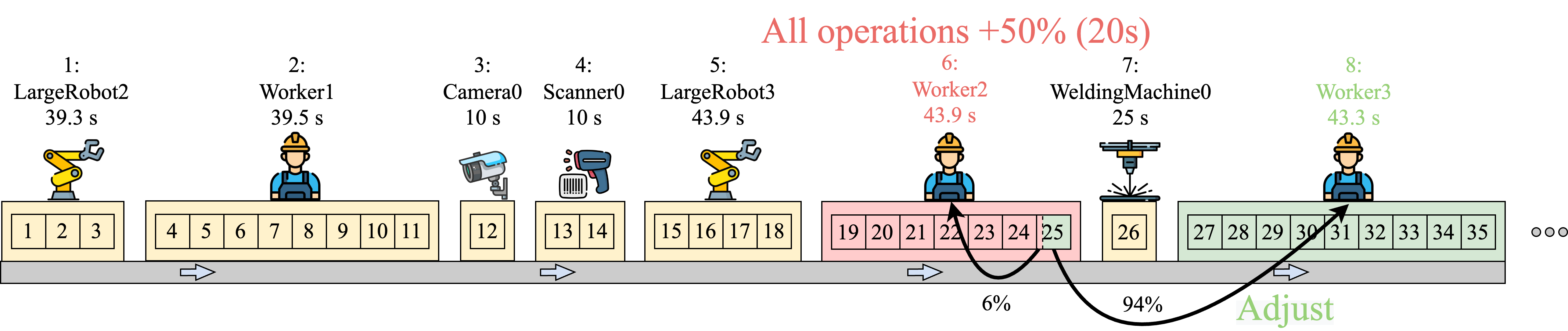}}
    \hfill
    \subfloat[\label{fig:line_disturbance_handle_addon}]{
    	\includegraphics[width=0.95\linewidth, trim=0 0 0 0, clip]{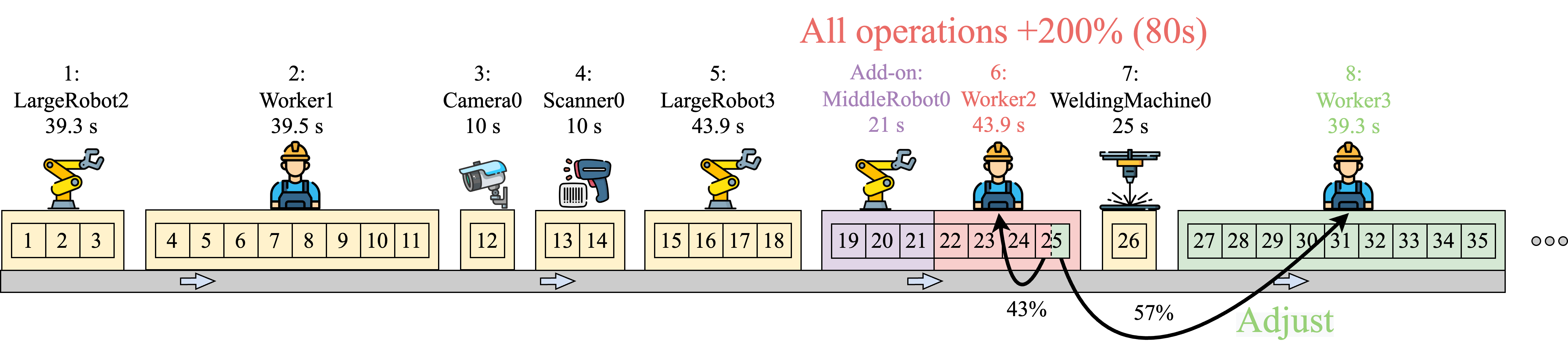}}
    \hfill
    \caption{An example manufacturing line operating under a disturbance due to a slower worker. There are 8 stations in the figure, with the numbers on the blocks indicating operations within a station performed by an agent.
    Reconfiguration is applied to maintain the bottleneck time. (a) Original manufacturing line configuration. (b) Redistributing tasks between the delayed worker to their neighbors. (c) Redistributing tasks between the delayed worker, a neighbor, and a newly added robot.}
    \label{fig:line_disturbance_handle}
\end{figure*}

\IEEEPARstart{T}he current manufacturing paradigm is shifting towards a more dynamic environment that includes product customization, demand fluctuation, and global collaboration~\cite{leitao2009agent,bi2023distributed}.
In addition to the variability introduced into the production pipeline, different sources of disturbances result in a delay of production, including worker absence, worker delay, and machine breakdown.
Consider the example manufacturing line in Fig. \ref{fig:line_disturbance_handle}. 
If Worker2 spends more time completing its operations, depending on the severity of the delay, there may become a need to initiate a line reconfiguration.
Throughput refers to the production rate, measured as the average number of products produced per hour. It is a key indicator of efficiency and productivity.
Fig. \ref{fig:line_disturbance_handle}b-c provide two examples of how to adjust the manufacturing line to maintain the desired throughput.

This reconfiguration scenario poses three important questions: What alternative agent options and line configurations are available in the system? What will be the impact on the performance (e.g., throughput) of the system due to the implementation of a reconfiguration option? How to determine which reconfiguration option is optimal for a given scenario?

Formally, suppose there is an ordered set of operations necessary to manufacture a product, \(J = \{1, 2, \cdots, |J|\}\), and a set of agents (workers/robots/machines), \(K = \{1, 2, \cdots, |K|\}\), each capable of performing a specific subset of the operations, \(J_k \subset J\). The symbol \(|\cdot|\) denotes the number of elements in a set. Initially, a plan is established to assign each required operation to a suitable agent, with the possibility of an agent handling multiple contiguous operations.
However, unforeseen disturbance events can occur, leading to some agents failing to achieve their standard throughput.
In response, the objective is to minimally adjust the existing assignment plan through a scoped reconfiguration and develop a new plan that optimizes overall throughput and the number of agents utilized, taking into account the disruptions.

To cope with the disturbances, reconfigurable manufacturing systems (RMSs) have been introduced to enable a flexible and dynamic response to unexpected disturbances~\cite{koren2010design}.
To guarantee reconfiguration performance, RMSs require the capabilities of real-time monitoring, interoperable system modeling, decision optimization, and automatic simulation for configuration evaluation~\cite{koren2018reconfigurable, yelles2021reconfigurable}.
However, the existing literature tends to focus narrowly on some aspects of these requirements and assume others are satisfied~\cite{koren2010design,koren2018reconfigurable,yelles2021reconfigurable}, such as proposing a digital twin framework for the entire system~\cite{leng2022digital} without considering the optimization and simulation for RMSs.
Few works study an integrated framework for RMSs, where the modules are developed using various approaches to combine the advantages, making the RMSs more flexible, responsive, and efficient for disturbance response.
Therefore, a digital twin-based dynamic line reconfiguration framework is developed in this paper to deal with this disturbance handling problem comprehensively.

\subsection{Summary of the Framework}
\begin{figure}[t!]
	\centering
	\includegraphics[width=0.7\linewidth]{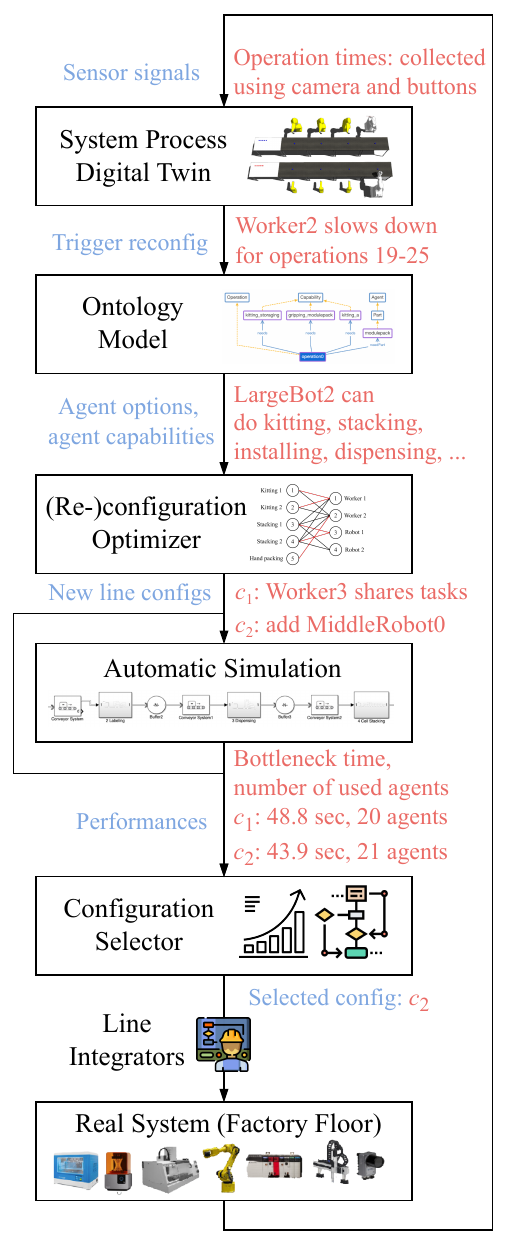}
	\caption{System diagram of the dynamic reconfiguration framework. The blue text indicates the input/output type. The red text shows examples used in this paper.}
	\label{fig:system_flowchart}
\end{figure}

This paper presents a framework to initialize and reconfigure manufacturing line structures through a combination of digital twins, ontology, discrete optimization, and automated simulation technologies. As shown in Fig. \ref{fig:system_flowchart}, the framework contains:
\begin{enumerate}[label={\arabic*)}]
    \item a system process digital twin (SDT) that monitors the real system and triggers a line reconfiguration according to the signals from the manufacturing line;
    \item an ontology model that organizes and stores resource options, line requirements, and practical constraints;
    \item a (re-)configuration optimizer that optimizes the expected performance and generates a set of Pareto optimal line configurations;
    \item a simulator that evaluates the set of optimal line configurations and conducts detailed performance analysis.  
\end{enumerate}

While the SDT and the ontology model are leveraged from existing work and applied to the case study in this paper, a (re-)configuration optimizer and an automatic simulation generation mechanism are proposed as innovative contributions to reconfigurable manufacturing systems.

\subsection{Contributions}
% The framework can be shown to be applied to real manufacturing lines.
This paper provides the following contributions.
\begin{enumerate}[label={\arabic*)}] %\setcounter{enumi}{-1}
    \item The development of a digital twin-based RMS to comprehensively achieve real-time monitoring, interoperable system modeling, decision optimization, and automatic simulation for high-performance reconfiguration of manufacturing lines.
    
    \item The formulation of a (re-)configuration optimizer that maintains system throughput by optimally redistributing the operations from a disrupted agent to neighboring or additional agents.
    \item The implementation of a simulation generation program that automatically generates discrete event simulation setups based on manufacturing line configurations and evaluates the system performance at around 400x real-time speed.
\end{enumerate}

\subsection{Outline}
Sec. \ref{sec:related_work} provides the related work about the digital twins, ontology, and mixed-integer programming and compares our work with existing reconfigurable manufacturing systems.
Sec. \ref{sec:method} describes the four major components in the dynamic reconfiguration framework.
Sec. \ref{sec:case_study} applies the reconfiguration framework to a practical case study to demonstrate the efficacy of the proposed method.
Sec. \ref{sec:conclusion} concludes the paper, summarizes the major results, and outlines future work to address.

\section{Related Work}\label{sec:related_work}

\newcommand{\reqsatisfied}{\(\bullet\)}
\newcommand{\requnsatisfied}{\(\circ\)}

\begin{table*}[tb]
\centering
\caption{Summary of requirements for dynamic line reconfiguration for disturbance handling}
\begin{tabular}{c c c c c c c c c}
\hline
\noalign{\vskip 6pt}
\multirow{2}{*}{Related Work}& Real-time  & \multirow{2}{*}{Interoperability}  & \multirow{2}{*}{Optimality}  & Process  & Mixed Human-  & Real  & Line & Dynamic Disturbance \\
& Monitoring  &   &   & Simulation  & Machine Factory  & Factory  & Reconfiguration & Handling\\
\hline
\cite{bi2021dynamic,bi2024dynamic} & \reqsatisfied{} & \requnsatisfied{} & \reqsatisfied{} & \reqsatisfied{} & \requnsatisfied{} & \requnsatisfied{} & \requnsatisfied{} & \reqsatisfied{}\\
\cite{poudel2022integrated} & \reqsatisfied{} & \reqsatisfied{} & \textbf{/} & \textbf{/} & \requnsatisfied{} & \requnsatisfied{} & \requnsatisfied{} & \textbf{/}\\
\cite{kovalenko2022toward} & \reqsatisfied{} & \requnsatisfied{} & \requnsatisfied{} & \requnsatisfied{} & \reqsatisfied{} & \reqsatisfied{} & \reqsatisfied{} & \reqsatisfied{}\\
\cite{yang2022intelligent} & \textbf{/} & \requnsatisfied{} & \requnsatisfied{} & \requnsatisfied{} & \requnsatisfied{} & \requnsatisfied{} & \reqsatisfied{} & \reqsatisfied{}\\
\cite{vahedi2022workforce} & \textbf{/} & \requnsatisfied{} & \reqsatisfied{} & \requnsatisfied{} & \reqsatisfied{} & \requnsatisfied{} & \requnsatisfied{} & \requnsatisfied{}\\
\cite{dou2020mixed,yelles2022minimizing,nakano2021manufacturing} & \textbf{/} & \requnsatisfied{} & \reqsatisfied{} & \requnsatisfied{} & \requnsatisfied{} & \requnsatisfied{} & \reqsatisfied{} & \requnsatisfied{}\\
\cite{bortolini2021optimisation} & \textbf{/} & \requnsatisfied{} & \reqsatisfied{} & \requnsatisfied{} & \requnsatisfied{} & \requnsatisfied{} & \textbf{/}& \requnsatisfied{}\\
\cite{qamsane2019unified} & \reqsatisfied{} & \requnsatisfied{} & \textbf{/} & \requnsatisfied{} & \requnsatisfied{} & \requnsatisfied{} & \requnsatisfied{} & \reqsatisfied{}\\
\cite{son2001automatic, martinez2018automatic} & \reqsatisfied{} & \requnsatisfied{} & \textbf{/} & \reqsatisfied{} & \requnsatisfied{} & \requnsatisfied{} & \reqsatisfied{} & \textbf{/}\\

\cite{delorme2024modelling} & \requnsatisfied{} & \requnsatisfied{} & \reqsatisfied{} & \requnsatisfied{} & \requnsatisfied{} & \requnsatisfied{} & \textbf{/} & \textbf{/}\\
\cite{kombaya2022digital} & \reqsatisfied{} & \reqsatisfied{} & \textbf{/} & \reqsatisfied{} & \requnsatisfied{} & \reqsatisfied{} & \reqsatisfied{} & \textbf{/}\\
\cite{arnarson2023towards} & \reqsatisfied{} & \requnsatisfied{} & \reqsatisfied{} & \reqsatisfied{} & \requnsatisfied{} & \requnsatisfied{} & \reqsatisfied{} & \requnsatisfied{}\\
\cite{caesar2023digital} & \reqsatisfied{} & \requnsatisfied{} & \reqsatisfied{} & \requnsatisfied{} & \requnsatisfied{} & \requnsatisfied{} & \reqsatisfied{} & \requnsatisfied{}\\
\cite{leng2020digital,leng2022digital} & \reqsatisfied{} & \reqsatisfied{} & \reqsatisfied{} & \reqsatisfied{} & \requnsatisfied{} & \requnsatisfied{} & \reqsatisfied{} & \requnsatisfied{}\\
\cite{zhang2019reconfigurable} & \reqsatisfied{} & \reqsatisfied{} & \textbf{/} & \requnsatisfied{} & \requnsatisfied{} & \requnsatisfied{} & \reqsatisfied{} & \requnsatisfied{}\\
\cite{mo2023framework} & \reqsatisfied{} & \reqsatisfied{} & \reqsatisfied{} & \reqsatisfied{} & \requnsatisfied{} & \reqsatisfied{} & \requnsatisfied{} & \requnsatisfied{}\\
\hline
This Work & \reqsatisfied{} & \reqsatisfied{} & \reqsatisfied{} & \reqsatisfied{} & \reqsatisfied{} & \reqsatisfied{} & \reqsatisfied{} & \reqsatisfied{}\\
\hline
\noalign{\vskip 3pt}

\noalign{\vskip 3pt}
\noalign{\vskip 3pt}
\multicolumn{9}{l}{\reqsatisfied{}\ and \requnsatisfied{}\ represent that the work does and does not satisfy the corresponding requirement, respectively;}\\
\multicolumn{9}{l}{ \textbf{/} represents that the work does not mention the corresponding requirement.}\\
\multicolumn{9}{l}{Combined references represent that they satisfy the same requirements under the categories in this table.}\\
\end{tabular}
\vspace{-12pt}
\label{tab:litreview}
\end{table*}

RMSs aim to quickly adjust their production capacity and functionality with limited resources in response to sudden changes in market or manufacturing systems~\cite{koren2010design, koren2018reconfigurable}.
Most existing RMS work focuses on reconfiguring the system based on the changing customer demand~\cite{pansare2023reconfigurable}, while the reallocation of the operations to handle disturbances to the system receives less attention.
Focusing on dynamic line reconfiguration for disturbance handling, we analyzed state-of-the-art research from the following requirements and functionalities: real-time monitoring, model interoperability, optimality, process simulation, mixed human-machine factory, real factory, line reconfiguration, and dynamic disturbance handling. \tableref{}~\ref{tab:litreview} reveals that none of the existing literature fully addresses all these requirements, with each work focusing on limited areas.
Our proposed framework, by encompassing all these dimensions, offers a holistic solution that enhances system adaptability, interoperability, and real-time responsiveness in dynamic manufacturing environments.

RMSs are supposed to promptly discover and identify disturbances in manufacturing systems, making real-time monitoring a necessary function.
Digital twin (DT), as a powerful tool for creating virtual replicas of physical systems and processes, has emerged and been applied for real-time monitoring and disturbance identification and prediction~\cite{kombaya2022digital,moyne2020requirements}.
However, many existing works applying DT to manufacturing systems only cover an introduction of DT with a simplified case study~\cite{qamsane2019unified, caesar2023digital}, not making full use of the advantages of DTs.
Other work proposes DT frameworks for the entire system\cite{kombaya2022digital,leng2020digital,leng2022digital}, requiring great efforts to design the DT.
Therefore, for designing RMSs, applying DT technology in a way that can balance the advantages and design efforts is necessary. 
In this work, we develop a DT of the system process for the purpose of real-time monitoring and providing information for simulation.

Once the disturbance is discovered, it is important to identify manufacturing resources that are available within the system for reconfiguration, thus we need a system-level capability model.
However, current capability models lack detailed information~\cite{qamsane2019unified} and/or have poor interoperability~\cite{bi2021dynamic, matsokis2010ontology}.
Therefore, researchers have begun to develop ontology techniques to model system capabilities.
Ontology is a formal, explicit specification of a shared abstract model to represent things in reality, as well as their properties, relationships, and restrictions using natural language~\cite{leitao2016industrial}.
In~\cite{jarvenpaa2019development, poudel2022integrated}, researchers use semantic descriptions and associated parameters to capture the resource capabilities, with additional hierarchical and co-operating relationships between resources for system capability modeling, which enables interoperability of information between different resources.
Nevertheless, these works do not explicitly introduce how the ontology model can be used for dynamic disturbance handling.
We leverage these capability-based ontology architectures and integrate the use of them into our reconfiguration framework to provide information regarding agent options and line constraints during the reconfiguration.

To obtain an optimal solution for manufacturing system reconfiguration, artificial intelligence-based methods have been investigated as a method for determining reconfiguration decisions, such as agent-based rescheduling~\cite{bi2021dynamic,bi2024dynamic}, automated learning control architectures~\cite{kovalenko2022toward}, and reinforcement learning~\cite{yang2022intelligent}.
However, these learning approaches cannot provide guarantees on the optimality of the reconfiguration decisions.
Thus, researchers have developed exact methods by formulating the decision-making problem as a mixed-integer linear program (MILP)~\cite{dou2020mixed,vahedi2022workforce,yelles2022minimizing,nakano2021manufacturing,bortolini2021optimisation}.
However, most existing literature only introduces an optimization model, lacking an integrated optimization-based framework for reconfiguration. In addition, none of these studies develop their optimization model for a mixed human-machine manufacturing line that is implemented in the real world.
In this work, we present a new MILP formulation for handling disturbances in a manufacturing line comprising machines, robots, and human workers. The MILP optimizer applies strategies for redistributing the manufacturing operations to existing agents as well as additional agents.

While the proposed optimization approach is capable of generating a set of reconfiguration options based on given manufacturing objectives, there is a need to evaluate these options for feasibility and expected performance within a simulated environment. 
Recent research has investigated approaches to automate the process of determining specific agent behaviors and updating system model parameters within a simulation environment \cite{son2001automatic,kovalenko2019model, martinez2018automatic, poudel2022integrated}. However, these simulations mainly focus on feasibility verification of the decision-making strategy rather than an evaluation of the system performance (e.g., throughput, operation time).
In this work, we demonstrate the development of a program to automatically generate discrete event simulations of manufacturing lines to evaluate system performance under varying disturbances. The agents (e.g. workers and robots) in the manufacturing line are updated based on a configuration file. The simulation considers several practical behaviors, including worker delay, worker absence, and task sharing between adjacent agents.

In summary, though a significant amount of work has studied the design and the components of RMSs, existing research lacks an integrated framework that connects real-time monitoring, system-level modeling, a practical optimization strategy, and an automated simulation environment for agile performance evaluation to enable efficient and dynamic reconfiguration for manufacturing lines.

\section{Dynamic Line Reconfiguration Framework}\label{sec:method}

This section presents the dynamic reconfiguration framework illustrated in Fig. \ref{fig:system_flowchart}. The four major components include a system process digital twin, an ontology model, a (re-) configuration optimizer, and a performance simulator that have been briefly introduced in Sec. \ref{sec:introduction} and will be detailed in this section. In this section, no assumptions have been made regarding what specific operations are included in the manufacturing line. Therefore, the framework can be generalized and applied to manufacturing lines consisting of operations to be done by agents, and we will conduct a case study in Sec. \ref{sec:case_study}.

\subsection{Key Concepts and Terminology}
This section defines the concepts that will be used throughout the paper.
\begin{enumerate}[label={\arabic*)}] %\setcounter{enumi}{-1}
    \item \textit{Agent:} An agent is an active unit that processes products in the manufacturing line. In this paper, it can either be a human worker, a robot, or a machine.
    \item \textit{Auxiliary resource:} An auxiliary resource is a resource used by an agent during the processing of the products. Examples include parts to be installed to the product, tools used by the agent to process the product, or storage bins for temporarily holding the parts and products.
    \item \textit{Operation:} A unit task performed by an agent. The time to perform this unit task is the operation time.
    \item \textit{Station:} A station consists of one or multiple operations, one agent (to perform the operations), and multiple auxiliary resources. The total time to perform all of the operations within the station is the station time.
    \item \textit{Bottleneck Time:} The maximum station time among all stations. In the example system from Fig. \ref{fig:line_disturbance_handle_baseline}, the bottleneck time is given as 43.9 s.
    \item \textit{Assignment plan:} An assignment plan matches an operation with an agent.
    \item \textit{Line configuration:} A plan for producing a part that includes a sequence of operations to be performed, a set of used agents, and an assignment.
    % A sequence of operations that includes a plan for the operation-agent assignments.
    \item \textit{Reconfiguration:} A change to the operation-agent assignment, including a plan switch or a configuration switch. In this paper, a reconfiguration does not consider changes to the operation sequence in a line.
    \item \textit{Plan Switch:} A reconfiguration in which the operation assignment is changed while the agents in the line remain the same. An example is shown in Fig. \ref{fig:line_disturbance_handle_adjust}.
    \item \textit{Configuration Switch:} A reconfiguration in which the operation assignment is changed while at least one agent is added or removed from the original manufacturing line. An example is shown in Fig. \ref{fig:line_disturbance_handle_addon}.
\end{enumerate}

\subsection{System Process Digital Twin}

In the reconfiguration process, the SDT takes a set of sensor signals, including camera inputs and button signals, evaluates them, and triggers reconfigurations, as illustrated in Fig. \ref{fig:system_flowchart}.

A digital twin is a purpose-driven dynamic digital replica of a physical asset, process, or system~\cite{moyne2020requirements}. 
In this paper, the system process digital twin is a digital twin of the processes in the manufacturing line system. It monitors the operational metrics of the manufacturing line and agents in the line for the purpose of system prescription and line reconfiguration.
Since the operations in the manufacturing line in this work are handled sequentially, a disturbance in a single agent invariably impacts the entire line.
Whenever an agent's metrics (e.g., throughput, operation times) deviate from the expected distribution stored in the ontology model and the deviation is larger than a preset threshold, the SDT marks the agent or station as experiencing a disturbance. The performance metrics of the agents are then utilized to determine if the disturbance is affecting the entire manufacturing line. If yes, the SDT initiates a reconfiguration process. This reconfiguration is designed to address the disturbance and restore the performance metrics. Below, we use an example to illustrate how our SDT can be applied in a manufacturing system. The performance metrics are defined, and the workflow is detailed.

Since throughput is the major metric to optimize in this paper, the SDT monitors the station times \(t_k\) of an agent (which is inversely proportional to the throughput) and the operation times \(T_{k j}\) for an agent \(k\) to perform \(j \in J_k\), and identifies time disturbances. Future work would extend the SDT to monitor quality or other important performance metrics and track other types of disturbances.

Consider the manufacturing system in Fig. \ref{fig:line_disturbance_handle_baseline} as an example. The figure shows a system with eight stations, where several operations are handled by a single agent within each station.
Leveraging the structural design of the SDT from \cite{poudel2022integrated, qamsane2021methodology}, the SDT monitors the time a product enters and departs a station to calculate the station time.
In a smart manufacturing station, the entering and departing of a product are tracked by cameras and photoelectric sensors installed above the conveyors, with robots sending out messages or humans pressing buttons before and after they perform a task on a part.

If the station time, \(t_k\), extends and deviates significantly from the distribution stored in the ontology model, leading to a reduction in the throughput, and this deviation persists beyond a predefined duration threshold, the SDT flags the operation in that specific station as disturbed and indicates the delay of agent \(k \in K\) as the cause. The SDT updates the operation time distribution \(T_{k j}\) of the agent (stored in the ontology model). If the agent loses the capability to perform an operation, the corresponding capability will be removed in the ontology model. Next, a signal is sent to trigger the reconfiguration decision-making process. The disturbances and agent performances are stochastic. To enable more robust estimation, agent operation times are modeled as random distributions and updated gradually when noisy samples are collected.

Anomaly classification methods \cite{toothman2023digital} should be applied to select an appropriate threshold that maximizes the accuracy of disturbance detection.
If the threshold value is too high, the system could miss a potential anomaly, resulting in significant production loss. However, selecting a smaller threshold will result in a more sensitive system, with the potential for several false positives and unnecessary system reconfigurations.

\subsection{Ontology Model}
During the reconfiguration process, an ontology model can efficiently retrieve a set of agent options and their associated capabilities based on the disturbance source (e.g., a specific worker operating at a slower pace).

Manufacturing line system-related data (e.g., agent capabilities and operation times) can be gathered and stored in a parsable file format, e.g., JSON format. However, data stored in such a format are not directly searchable for planning and archiving purposes. The distinct data format from different subsystems also makes data utilization challenging. In this work, we leverage a capability-based ontology model \cite{wan2018ontology,jarvenpaa2019development,poudel2022integrated} as a flexible and dynamic way to encode objects and properties from a manufacturing line. The model generalizes to different types of manufacturing lines, standardizes the representation of the manufacturing data, and, therefore, facilitates efficient information retrieval for applications like configuration optimization.

There are two components used in an ontology model: objects and properties. Objects include agents (e.g., workers, robots), resources (e.g., parts, storage bins, tools), capabilities, operations, stations, and manufacturing lines.
Properties describe details of an object (data properties) or the relationship between two objects (object properties).
Data properties describe the inherent details of an object, such as the identification number (ID) or the type of robot. Object properties describe the relationship between two object entities. For instance, a LargeRobot \highlightontology{has} a kitting capability, indicating the robot is able to perform a kitting operation with the correct auxiliary agents. For object properties, the relationships between the objects in the ontology may be represented through a graph.
Objects and properties serve as nodes and edges to form an ontology graph that stores complex information.

\begin{figure}[t!]
    \centering
    \includegraphics[width=1.0\linewidth]{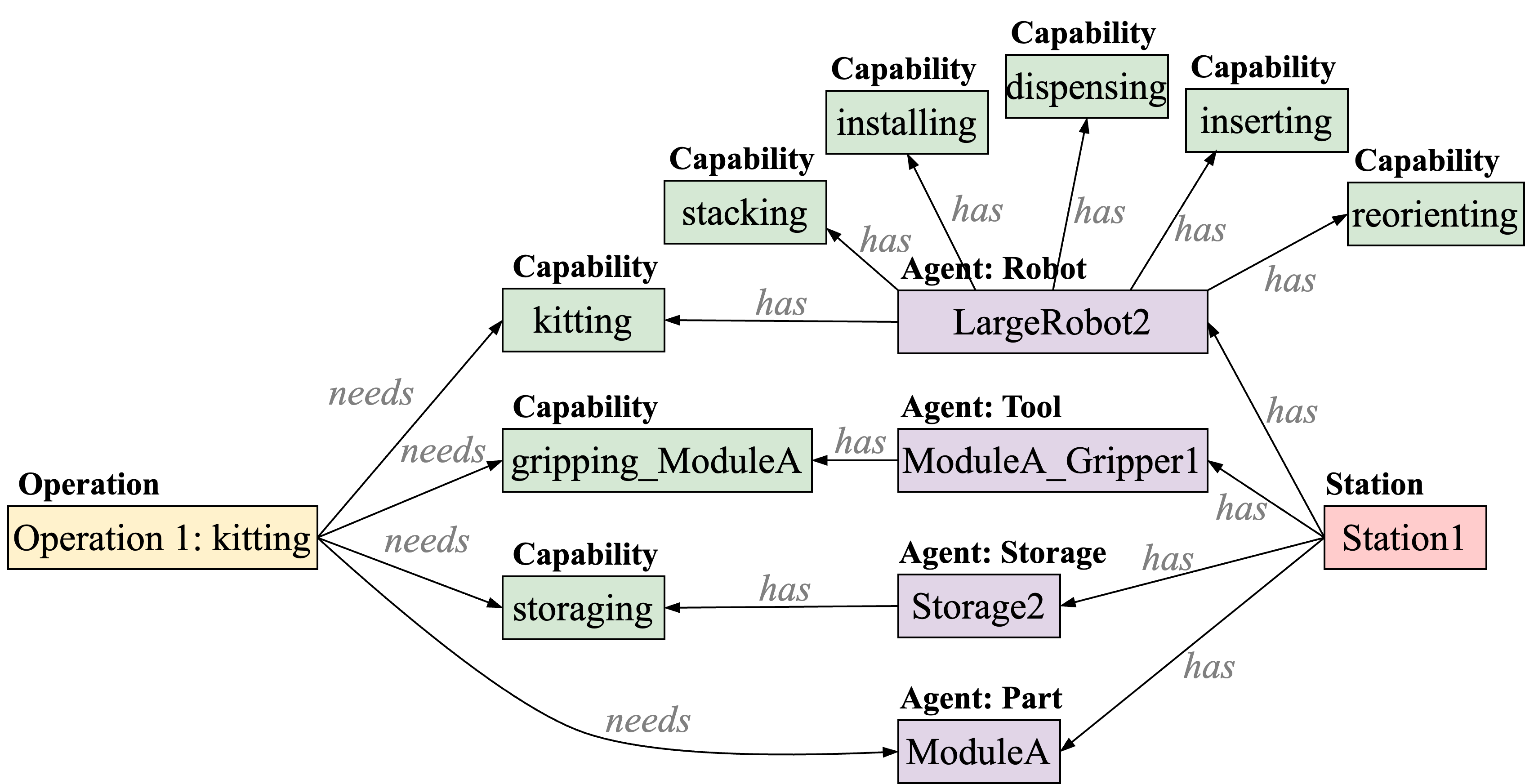} 
    \caption{A part of the ontology graph. An operation needs several capabilities and parts to complete. A station contains agents to fulfill the needs of an operation.}
    \label{fig:ontology_example}
\end{figure}

An ontology model is typically constructed based on resource descriptions and process requirements from manufacturers. For example, a LargeRobot with kitting, stacking, and installing end-effector tools is registered with kitting, stacking, and installing capabilities. In this work, the capability registration is performed automatically based on the description sheets of the agents and resources. A station consists of multiple objects. If a station comprising a robot, a storage resource, and a gripper to process a ModuleA part, the ontology graph would then be defined to describe that Station1 \highlightontology{contains} these objects, as shown in Fig.~\ref{fig:ontology_example}.
In addition, resource capabilities can be transferred. Since the LargeRobot \highlightontology{has} the kitting capability, the station also \highlightontology{has} the kitting capability.
Considering the needed capabilities and parts of Operation1 (a kitting operation), Station1 \highlightontology{has} all the \highlightontology{needs} to perform Operation1.
Instead of relying on a massive list of resource descriptions, the ontology model organizes all resource descriptions and process requirements into a holistic graph represented in natural language. This approach makes the manufacturing system model modular, interoperable, and user-friendly.

Through the ontology model, once the sequence of operations in a manufacturing line has been determined, it is straightforward for users to identify agents and their capabilities needed to perform the necessary operations (as in Fig. \ref{fig:ontology_example}).
Recalling the operation set \(J = \{1, 2, \cdots, |J|\}\) to be performed for a product and the available agent set \(K = \{1, 2, \cdots, |K|\}\), ontology queries identify the relationship between agents and operations, expressed as an achievable operation set \(J_k \subset J\) for agent \(k \in K\). In addition, it extracts the operation time \(T_{k j}\) for agent \(k \in K\) to perform operation \(j \in J_k\).
This resource information is then sent to the configuration optimizer.

As part of this framework, we developed a program to read manufacturing data and automatically generate an ontology graph stored in an XML file. The XML file can be visualized using existing software (e.g., Protégé \cite{musen2015protege}) and searched efficiently using an ontology SPARQL query\footnote{SPARQL is a query language for RDF (Resource Description Framework) that enables retrieval and manipulation of data stored in the RDF format.}.
Note that the ontology model in this paper aligns with the Semantic Web standards developed by the World Wide Web Consortium (W3C)~\cite{trappey2016review} and ISO/IEC 21831~\cite{iso21838}.

\subsection{Configuration Optimizer}\label{sec:line_config_optimizer}
The configuration optimizer gathers agent options and capabilities from the ontology model and generates a list of potential line configurations aimed at mitigating the disturbance and restoring the manufacturing throughput.

In this subsection, we formulate an innovative optimization problem to encode the practical constraints of line reconfiguration.
The inputs of the optimizer are the available agent options, agents' capability to perform operations, and constraints about the scope of the optimization according to the disturbance. The output is an assignment plan that matches each operation with one agent and minimizes the bottleneck time with a limited number of agents. The optimizer is guaranteed to generate solutions if enough agents are provided to complete all the operations. If no solution is generated, the line integrators need to check if more agents are available and add them to the ontology model.

The operations in a manufacturing line have to be performed in a fixed sequence with agents assigned to each task.
Consider a line consisting of a sequence of fixed operations that must be completed by agents (workers/robots/machines). Let the set of operations be \(J = \{1, 2, \cdots, |J|\}\), and the set of agents be \(K = \{1, 2, \cdots, |K|\}\), where \(|\cdot|\) denotes the number of elements in a set.
Let \(J_k \subset J\) be the set of operations that an agent is able to perform (this information is collected from the ontology).
An agent \(k \in K\) is allowed to perform multiple contiguous operations.
Each operation \(j \in J\) should be assigned to precisely one agent \(k \in K\) provided that \(j \in J_k\).

\begin{figure}[t!]
    \centering
    \includegraphics[width=0.7\linewidth]{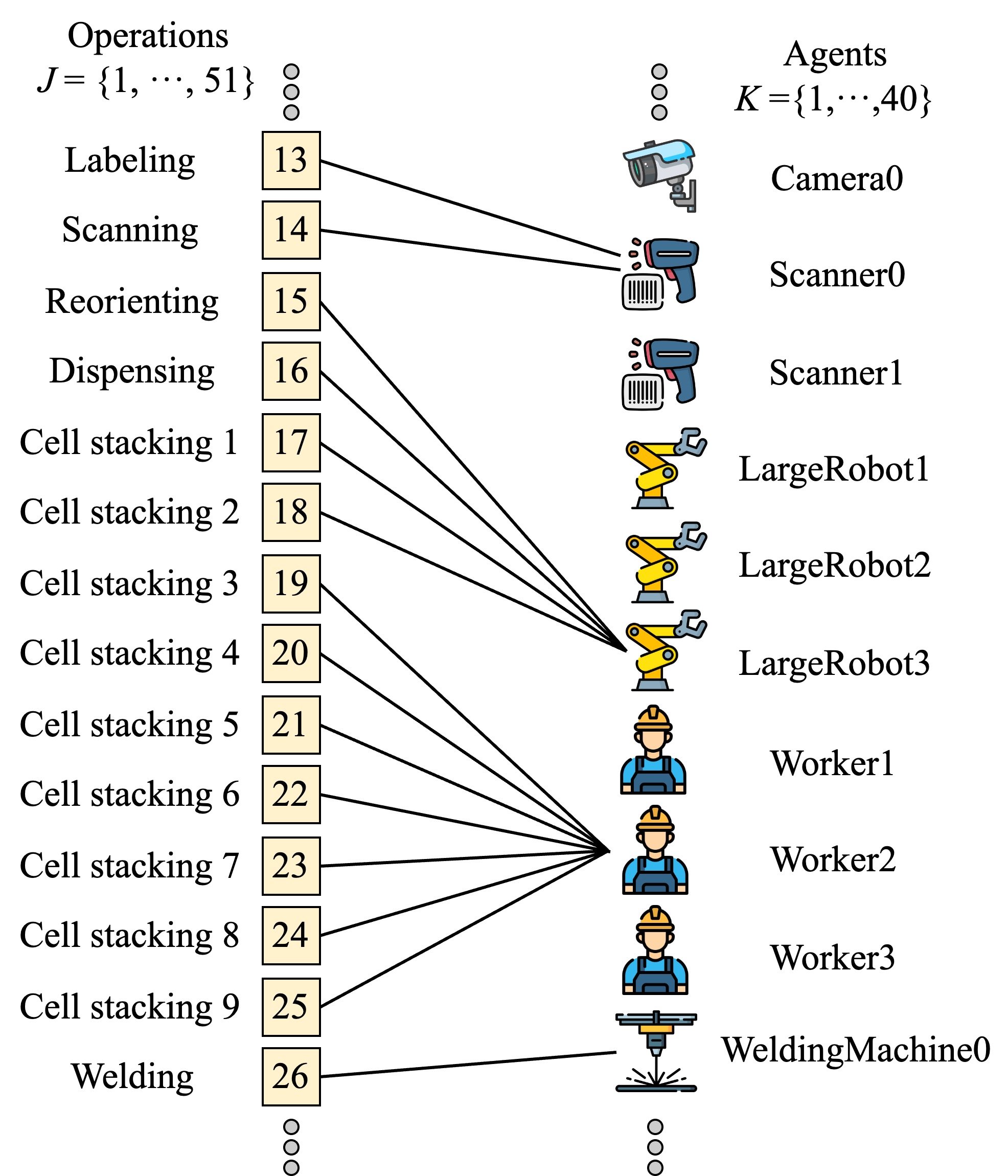}
    \caption{A graphical model for a matching problem with 51 operations and 40 available agents deriving from 8 agent types.
    The edges indicate the agents selected for a particular task.
    In this example, Scanner0 performs the labeling and scanning operations, LargeRobot3 and Worker2 together perform the stacking operations, and WeldingMachine0 performs the welding operation.}
    \label{fig:operation_agent_matching}
\end{figure}

An optimization of the manufacturing production line determines which agent should perform a particular operation such that the bottleneck time is minimized given a limited number of available agents.
Since throughput is inversely proportional to the bottleneck time, this optimization algorithm aims at maximizing the throughput of the system.
Consider that agent \(k\) spends a total time of \(t_k\) on its assigned operations. If this is the bottleneck time, this time will be the maximum \(t_k\) over all agents.
The bottleneck time indicates how fast a product can be manufactured.
Graphically, the algorithm searches to find the optimal matching (Fig. \ref{fig:operation_agent_matching}) between the operations and agents, subject to practical constraints. Note that an edge can only exist if the agent can perform the corresponding operation. For instance, for the edges selected in Fig. \ref{fig:operation_agent_matching}, the corresponding manufacturing line configuration is shown in Fig. \ref{fig:line_disturbance_handle_baseline}. In this line setup, the bottleneck time is shown to be 43.9 s.

The configuration optimizer performs two functionalities: configuration initialization and reconfiguration. During the configuration initialization, we assume agent assignments are optimized across the entire manufacturing line. During reconfiguration, we preserve as much of the original manufacturing line configuration as possible, while optimizing the agent assignments around the disrupted agent.

\subsubsection{Line Configuration Initialization}\label{sec:line_config_initialization}
A mixed-integer program is formulated based on the graphical model in Fig. \ref{fig:operation_agent_matching}. Common notations are explained in Table \ref{tab:variable_definition}. Note that the decision variables are \(x_{k j}\), \(y_{k j}\), \(z_{k}\), and \(t_{k}\).

\begin{table}[h]
% \normalsize
  \caption{Definition of the notations for the optimization.}
  \label{tab:variable_definition}%
    \begin{tabular}{p{0.06\linewidth}|p{0.8\linewidth}} 
    \toprule
     & Meaning
    \\
    \midrule
    \(x_{k j}\)& = 1 if agent \(k \in K\) is assigned to perform operation \(j \in J\).
    \\
    \(y_{k j}\)& a helper variable for expressing the relationships between \(x_{k j}\) and \(z_{k}\) in \eqref{eqn:y_constraint}-\eqref{eqn:agent_usage_constraint}.
    \\
    \(z_{k}\) & the number of \textit{\textbf{contiguous}} usages of an agent. For example, if an agent is assigned to operations 1, 2, 3, 4, 5, 6, the usage value is 1, while an agent assigned to operations 1, 2, 3, 5, 6, has a usage value of 2. 
    \\
    \(t_{k}\) & the time for an agent to complete all its assigned operations.
    \\
    \(T_{k j}\) & the time that agent \(k \in K\) will spend at operation \(j \in J_k\). This is a stochastic parameter modeled as a random distribution.
    \\
    \bottomrule
    \end{tabular}
\end{table}

\textit{Objective Function:}
The objective function minimizes a weighted sum of the bottleneck time and the number of agents used, where \(C_t\) and \(C_z\) are weights on the time and number of agents, respectively. Since bottleneck time is a stochastic value, we add an expectation, \(E(t_k)\), in the objective function.
\begin{align}
    \min_{} \left\{ C_t \max_{k \in K}{E(t_k)} + C_z \sum_{k \in K} z_k \right\}\label{eqn:line_config_objective}
\end{align}

\textit{Variable Bounds:}
\(x\), \(y\) variables are binary, \(z\) variables are non-negative integers, while \(t\) variables are continuous.
\begin{align}
    x_{kj}, y_{kj} \in \{0,1\}, \ z_{k} \in \Zset{}, \ t_{k} \geq 0, \ \forall j \in J, \ \forall k \in K.\label{eqn:bound_constraint}
\end{align}

\textit{Time constraints:}
The time for an agent to complete all its assigned operations is the sum of the operation times.
\begin{align}
    t_k = \sum_{j \in J} {T_{k j} \cdot x_{k j} }, \quad \forall k \in K. \label{eqn:time_constraint}
\end{align}

\textit{Assignment constraints:}
Each operation must be assigned to exactly one agent.
\begin{align}
    1 = \sum_{k \in K} {x_{k j}}, \quad \forall j \in J. \label{eqn:assignment_constraint}
\end{align}

\textit{Capability constraints:}
An agent can only be assigned to operations for which it has the capability to perform.
\begin{align}
    0 = x_{k j}, \quad \forall j \notin J_k, \ \forall k \in K. \label{eqn:capability_constraint}
\end{align}

\textit{Agent usage constraints:}
\(y_{k j}\) equals 1 when an agent \(k \in K\) performs operation \(j\) but not \(j - 1\). Therefore, the summation of \(y_{k j}\) indicates the number of contiguous usages of an agent. For simplicity, we limit this usage to one in the configuration initialization problem; i.e., an agent can only be allowed to perform a contiguous subset of the operations.
% This agent, together with the assigned operations, becomes a station in the manufacturing line.
\begin{align}
    y_{k j} &= x_{k j}, \quad\quad\quad\quad\quad \ j=1, \quad\quad\quad \ \ \ \forall k \in K, \label{eqn:y_constraint} \\
    y_{k j} &\geq x_{k j} - x_{k (j-1)}, \quad j=2,\cdots,|J|, \ \forall k \in K, \\
    z_k &= \sum_{j \in J} {y_{k j}} \leq 1, \quad\quad\quad\quad\quad\quad\quad\quad\ \ \forall k \in K. \label{eqn:agent_usage_constraint}
\end{align}

% \textit{Agent priority constraints (optional):}
% If two agents are the same type, then the one with the smaller index will be used first. This is an optional symmetry-breaking constraint to reduce the number of solutions with the same objective value.
% Suppose the type of agent \(k \in K\) is denoted as \(a_k \in A = \{1,\cdots,|A|\}\). Then, this constraint can be mathematically presented as follows
% \begin{align}
%     z_{k_1} \geq z_{k_2}, \quad \forall k_1, k_2 \in K \text{ s.t. } k_1 \leq k_2, a_{k_1} = a_{k_2}. 
% \end{align}

\subsubsection{Line Reconfiguration}\label{sec:line_reconfig} Suppose we are given a manufacturing line setup in which we know all initial operation/agent assignments. In the event that there is a disturbance, and one of the agents begins performing much slower than normal, we only want to modify the neighboring operation/agent assignments rather than globally re-optimize the entire line configuration. Consider the example in Fig. \ref{fig:line_disturbance_handle}. The original line setup is shown in Fig. \ref{fig:line_disturbance_handle_baseline}.
If a disruption occurs by a worker becoming slower, a simple and low-cost strategy is to share the impacted worker's tasks with adjacent agents with the same capability. If the disturbance is large enough, additional agents can be incorporated into the line to handle the delayed operations.

During a reconfiguration, the disrupted agent and its neighboring agents are allowed to perform multiple non-contiguous operations. Furthermore, an agent is allowed to perform part of an operation.
We will discuss the feasibility and analyze the agent behaviors introduced by these assumptions in Sec. \ref{sec:case_study}.
With this relaxation, manufacturing line reconfiguration can introduce more flexible pairing options (e.g. task sharing between agents).
To include these partial assignment options, we add a continuous variable \(\Delta x_{k j}\) to indicate the adjustment to the binary task assignment \(x_{k j}\).

Let the operations originally performed by the disrupted agent be contained within set \(J_D\). Suppose the disturbed agent, the adjacent agents, other used agents in the line, and unused agents are in set \(K_D\), \(K_A\), \(K_L\), and \(K_U\), respectively.
For the example in Fig. \ref{fig:line_disturbance_handle}b-c, \(J_D = \{19, 20, \cdots, 25\}\), \(K_D = \{6\}\), \(K_A=\{5, 7, 8\}\), \(K_L = \{1,2,3,4\}\), and \(K_U = \{ \text{The IDs of any agent that is not in Fig. \ref{fig:line_disturbance_handle_baseline}}\}\).
Denote the solution corresponding to an original line configuration as \(x_{k j}^0\).
Then the reconfiguration can be described as the following optimization.
\begin{align}
    & \min_{} \ C_t \max_{k \in K}{E(t_k)} + C_z \sum_{k \in K } z_k \nonumber \\
    & \quad + C_x \sum_{k \in K, j \in J} |x_{k j} + \Delta x_{k j} - x_{k j}^0| \label{eqn:reconfig_objective} \\
    & \text{s.t. } \text{ \eqref{eqn:capability_constraint}-\eqref{eqn:agent_usage_constraint} and} \nonumber \\ 
    & \quad \Delta x_{k j} \in \begin{cases} [-1, 1], \quad \ \ \forall j \in J_k, \ k \in K_D\cup K_A
    \\\{0\},  \quad\quad\quad \text{otherwise}
    \end{cases}\label{eqn:deltax_bound}
\end{align}
\begin{align}
    & \quad 0 \leq x_{k j} + \Delta x_{k j} \leq 1, \quad\quad\quad \ \forall j \in J, \ \forall k \in K \label{eqn:reconfig_assignment_bound} \\
    & \quad t_k = \sum_{j \in J} {T_{k j} \cdot {(x_{k j} + \Delta x_{k j}) } },  \quad\quad\quad\quad \forall k \in K. \label{eqn:reconfig_time_constraint} \\
    & \quad 1 = \sum_{k \in K} {(x_{k j} + \Delta x_{k j})}, \quad\quad\ \ \forall j \in J. \label{eqn:reconfig_assignment_constraint} \\
    & \quad x_{k j} = x_{k j}^0, \quad\quad\quad\quad\quad\quad\quad\quad \forall j \notin J_D, \ \forall k \in K. \label{eqn:reconfig_origin_config}
\end{align}

According to \eqref{eqn:deltax_bound}, the adjustment variable \(\Delta x_{k j}\) can be nonzero only when \(j\) is a disturbed operation, and \(k\) is either the disturbed or adjacent agent. In addition, the agent assignments \(x_{k j}\) in \eqref{eqn:time_constraint}-\eqref{eqn:assignment_constraint} are replaced with \(x_{k j} + \Delta x_{k j}\) in \eqref{eqn:reconfig_time_constraint}-\eqref{eqn:reconfig_assignment_constraint}, while \(x_{k j} + \Delta x_{k j}\) should still be bounded between 0 and 1 as given in \eqref{eqn:reconfig_assignment_bound}.
Equation \eqref{eqn:reconfig_origin_config} ensures the assignment for the undisturbed operations remains unchanged by restricting the assignment variables of the unrelated operations and agents to their original values.
Finally, any adjustment to the original line configuration should be penalized as in \eqref{eqn:reconfig_objective}.
Note that the weight of the additional penalty associated with plan adjustments, \(C_x\), should be significantly smaller than \(C_z\) and \(C_t\).
This ensures that it avoids unnecessary adjustments to the original plan while maintaining the optimization goal of reducing bottleneck time and the number of used agents.
With the solutions to the assignment variables \(x_{k j}\) and the adjustment variables \(\Delta x_{k j}\), we can retrieve a new assignment plan to handle the disturbances.

By varying the weights \(C_t\) and \(C_z\), the optimization can generate a set of different line configurations to tackle disturbances. These configurations can be categorized into two major modes.
Plan switch: when more penalty is placed on the number of used agents, the optimizer tends to handle the disturbance by sharing tasks with neighbors.
Configuration switch: when more penalty is placed on bottleneck time, the optimizer handles the disturbance through both sharing tasks and adding additional agents into the line. The switch between the two modes is made smoothly by varying the weights.

\subsection{Automatic Performance Simulation}
As illustrated in Fig. \ref{fig:system_flowchart}, the component following the reconfiguration optimizer is the simulation generation program. This program automatically generates simulation models based on the new line configurations provided by the optimizer to evaluate system performance, including bottleneck time and the number of used agents.

Conventionally, simulations for manufacturing lines are built manually using platforms including Tecnomatix and SimEvents. However, for dynamic reconfiguration, new simulation setups should be automatically generated according to the line configurations from the configuration optimizer.
While existing work for automatic simulation focuses more on model parameter estimation and feasibility verification of a new line setup \cite{son2001automatic,kovalenko2019model, martinez2018automatic}, we develop a simulation generation program to automatically evaluate line performance (e.g., throughput) considering worker delay, worker absence, the breakdown of machines, and task sharing between agents.

The simulation program reads the configuration files detailing the ordered flow of stations in the manufacturing line, the agent and operation assignments in each station (given by the solutions \(x_{k j}\) from the optimizer), and the buffer locations and capacities (from the ontology model).
Then, using the interfaces provided by Mathworks, a MATLAB program automatically generates a simulation model that runs with SimEvents. SimEvent is a discrete event simulation environment where a timestamp is only added when an event occurs (e.g., a part departing a station or a worker completing an operation).
Each operation in the manufacturing line is a module in the SimEvent model and each agent is associated with a set of operations. The model simulates the part-processing workflow and operation durations while abstracting physical motions. As a result of its event-based design and the simplification of physical movements, the system runs hundreds of times faster than real time.

The simulation leverages the estimated operation time distributions, \(T_{k j}\), based on the data from the real manufacturing line (stored in the ontology model).
Then, the simulation sample operation times for each agent as the parts flow through the manufacturing line. The simulation generates key performance indicators such as throughput, cycle time, operation time, and buffer usage. Note that since the configuration optimizer algorithm in the previous subsection optimizes across average performances, the simulation program provides additional performance details for each configuration choice.

Finally, a MATLAB-based visualization program can generate an animation of selected periods during the simulation to illustrate important aspects of the simulation progress, such as parts flowing through the stations or parts getting stuck and blocking the manufacturing line.

\subsection{Configuration Selector}
The last component in the decision-making framework is a configuration selector that selects a single reconfiguration plan based on the simulated performance and a weighted objective.

The selector includes a set of user-defined performance thresholds to check the feasibility of the configurations. Within the set of feasible configurations, an optimal reconfiguration plan can be determined based on a user-defined objective.
For example, in the case study in Sec. \ref{sec:case_study}, the throughput of a manufacturing line is listed as the main objective. If there are multiple line reconfiguration plans that produce the same throughput, then the selector chooses the line plan that uses the fewest agents.
Compared to the configuration optimizer, the selector uses more detailed performance metrics that become available through the simulation environment to determine the optimal plan.

\section{Case Study}\label{sec:case_study}

\begin{figure}[t!]
	\centering
	\includegraphics[width=0.8\linewidth]{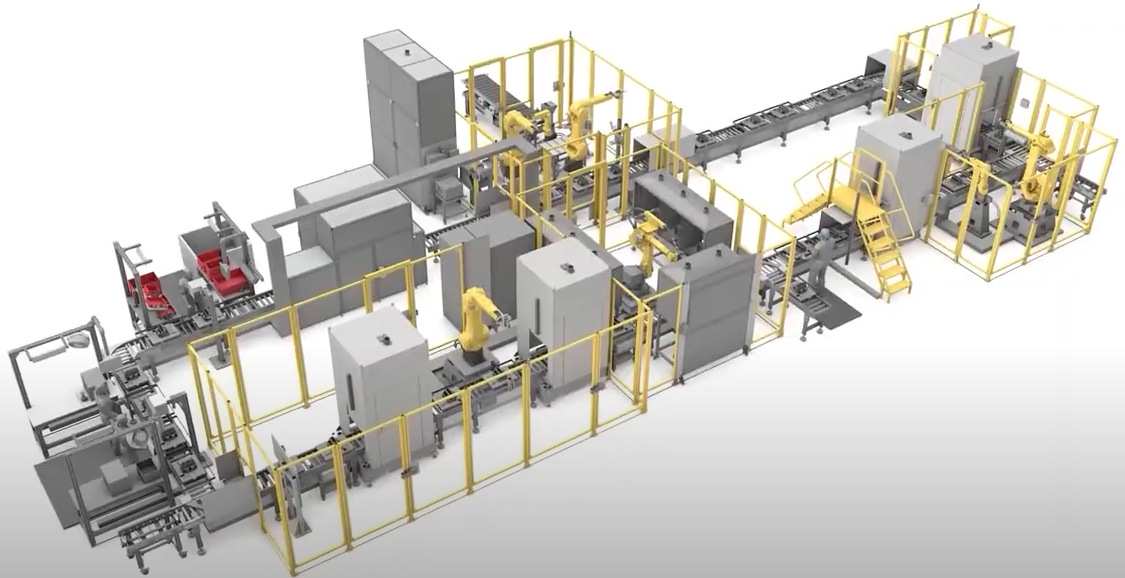}
	\caption{A battery manufacturing line.}
	\label{fig:real_manufacturing_line}
\end{figure}

In this section, we demonstrate the efficacy of the proposed dynamic reconfiguration framework using data and manufacturing line plans from a real battery manufacturing line, shown in Fig. \ref{fig:real_manufacturing_line}. The simulation environment is built based on the real manufacturing line.
Although the framework is demonstrated using this example manufacturing system, similar frameworks with the digital twin, ontology model, configuration optimizer, and automatic simulation strategy can be applied in other forms of manufacturing lines.
The manufacturing line consists of 51 operations to be performed by 8 different agent types with the appropriate capabilities. These agent types include human workers, large/middle/small-sized robots, camera inspectors, scanning machines, welding machines, and leak testers.
As mentioned in Sec. \ref{sec:method}, agents are modeled as objects in the ontology model, with associated properties, including capabilities. In the automatic SimEvent simulation, each agent is associated with a set of operation modules.
Examples of the operations performed by these agents include kitting, cell stacking, component installation, welding, and inspection. For each operation, there exist auxiliary resources, such as parts to be installed, storage tools, and grippers if the operation is handled by a robot.
We will step through the process described in Fig. \ref{fig:system_flowchart} to show how to apply the (re)configuration framework to initialize the flow within a manufacturing line and demonstrate how to conduct a reconfiguration to handle disturbances.
Although the case study focuses on line reconfiguration within a battery manufacturing line, the methodology presented can be effectively applied across a diverse range of manufacturing lines that share common elements such as workers, robots, machines, and auxiliary tools.

In the case study, our simulation leverages real manufacturing data provided by Hitachi to simulate operation times. For example, Fig. \ref{fig:normal_worker_operation_time} shows the time distribution for a worker to complete a cell stacking operation.
The computations for the optimization and simulation are done on a laptop with an Apple M1 chip.

\subsection{System Process Digital Twin}

The system process digital twin of the battery manufacturing line illustrated in Fig. \ref{fig:real_manufacturing_line} contains a manufacturing line monitoring system developed by Hitachi for data gathering and MATLAB-based models of the manufacturing line and agents for efficient updates. 
%closely monitor the time spent at each station and detect if a disturbance happens.
The methodology for building digital twins is based on our previous work~\cite{qamsane2021methodology} and aligns with the industry standard ISO 23247 (Digital Twin Manufacturing Framework)~\cite{iso2020automation}.

\begin{figure}[t!]
	\centering
	\includegraphics[width=\linewidth]{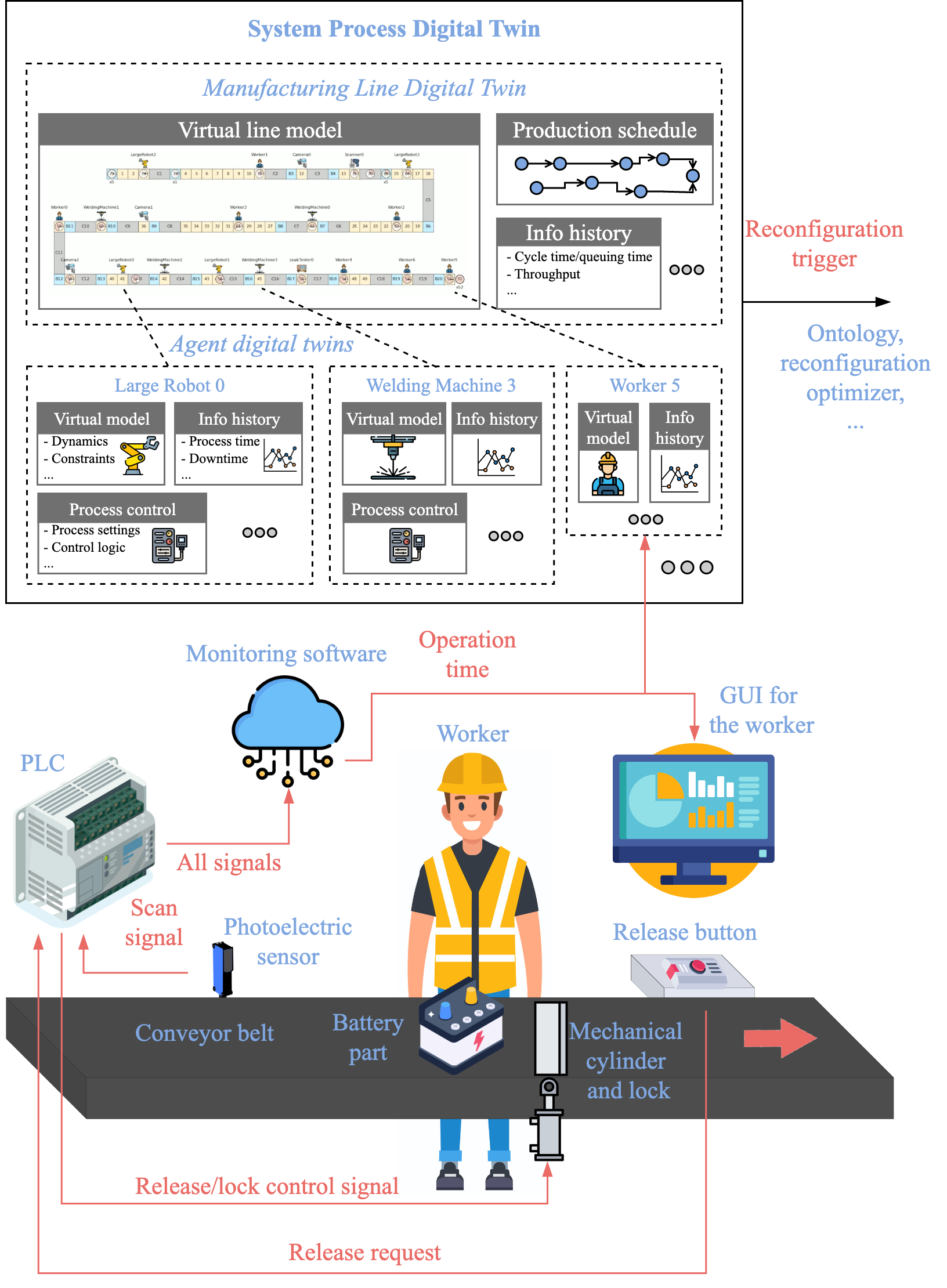}
	\caption{An illustration of the system process digital twin and its connection with the real manufacturing line.}
	\label{fig:system_digital_twin}
\end{figure}

As illustrated in Fig. \ref{fig:system_digital_twin}, photoelectric sensors have been installed on the conveyor to detect the arrival of a product. Once a product arrives, a sensor signal is sent to the programmable logic controller (PLC). Then, the PLC sends a signal to lock the product with a mechanical cylinder and starts a timer for the corresponding operations. After the completion of the assigned operations, the worker pushes a button to unlock the product. This button push signal is sent to the PLC, and the operation timer stops to return the time measurement. Note that if the agent for the operation is a robot or machine, then instead of pressing a release button, they directly send a release request to the PLC to release the part after the completion.
The delay in the collection of operation time (enabled by the photoelectric sensor, release button, and robot/machine signals) is negligible (milliseconds) compared to the operation time.

A manufacturing line monitoring software connected to the PLC stores the data from the timer, shows an average operation time for each worker and robot, and updates the operation time of workers in the agent digital twins.

The system process digital twin contains the manufacturing line digital twin and the agent digital twins in the line.
The agent digital twins are connected to an ontology model to retrieve information on their capabilities. Operation times from the monitoring software are incorporated into the agent digital twin. This capability data and operation time are then transferred to the manufacturing line digital twin. Specifically, in this case study, the human DT was modeled based on the historical data of the workers’ capabilities and performance, including skills and processing time.
Note that DT allows model aggregation~\cite{moyne2020requirements}, if additional data could be collected, such as fatigue, it can also be implemented in the human DT.
If a disturbance causes the operation time of an agent (a worker, robot, or machine) to deviate from the expected distribution stored in the ontology model, leading to reduced throughput, the operation time and the capability of the disturbed agent in the ontology model is updated, and the reconfiguration process is initiated.

\subsection{Line Configuration Initialization}
\begin{figure}[t!]
	\centering
	\includegraphics[width=1.0\linewidth]{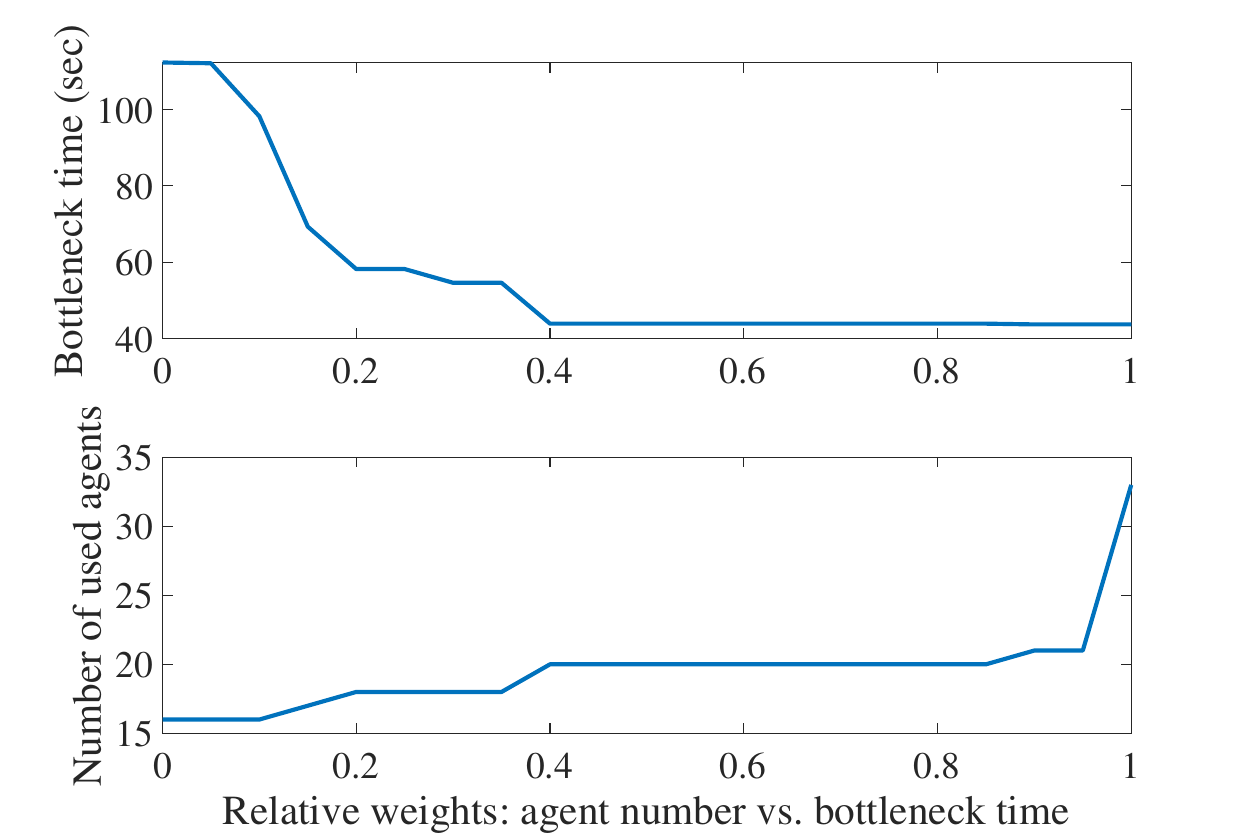}
	\caption{Trade-off between bottleneck time and the number of used agents.}
	\label{fig:initial_tradeoff_plot}
\end{figure}

Given the time distribution for an agent to perform an operation, we want to minimize the bottleneck time while limiting the number of agents used in the manufacturing line. This goal is achieved by applying the configuration initialization optimization described in Sec. \ref{sec:line_config_optimizer}. By changing the weights in the objective function \eqref{eqn:line_config_objective}, we first generate a set of optimal initial line configurations. A trade-off between the bottleneck time and number of agents used in the system can be seen in  Fig. \ref{fig:initial_tradeoff_plot}. As the weight on the bottleneck time, \(C_t\), is increased from 0 to 1 (recall that \(C_z = 1 - C_t\)), we observe that the minimum bottleneck time of 43.9 seconds is achieved for \(C_t\) values of 0.4-0.85 with the use of 20 agents.

\begin{figure}[t!]
	\centering
	\includegraphics[width=1.0\linewidth]{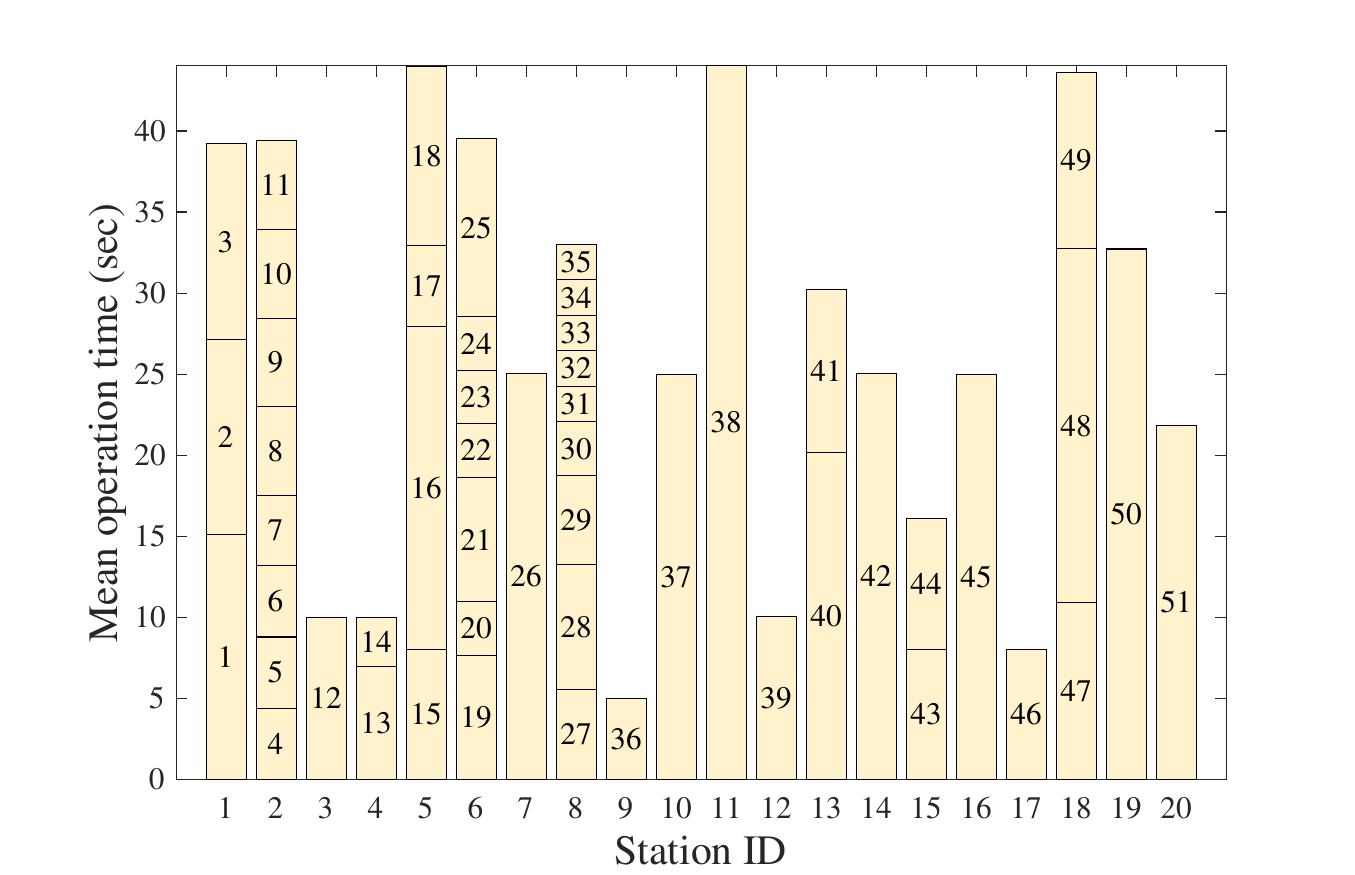}
	\caption{Initial configuration: The mean station and operation times according to the simulation. Each bar shows the total operation time of a station, while the smaller blocks within a bar indicate the operations within a station.}
	\label{fig:initial_station_operation_time}
\end{figure}

From the output optimizer, there are two feasible configurations when \(C_t\) is set to a value between 0.4-0.85 and the system uses 20 agents. 
Using the simulation program, we automatically generate the setups based on these two line configurations and evaluate their performances. The results indicate that one of the setups has more balanced station operation times, fewer bottlenecks in the flow line, and a slightly larger throughput. Thus, this configuration is selected as the initial starting point for the manufacturing production plan. 
The first eight stations in this initial manufacturing line are illustrated in Fig. \ref{fig:line_disturbance_handle_baseline}. The operation and station times are shown in Fig. \ref{fig:initial_station_operation_time}, where stations 5, 11, and 18 can be identified as the three bottlenecks in the production line. Based on the operation times, a new product can be generated on average every 43.9 s. Thus, the throughput in a 16-hour time slot is 1295 products. If we dig a bit deeper into Fig. \ref{fig:line_disturbance_handle_baseline}, we can see that operation 38 requires an average operation time of 43.9 s, and hence, the bottleneck time for any line configuration must be at least 43.9 s to accommodate this operation. %Therefore, the current bottleneck time is indeed optimal.

To evaluate the computational cost of the initial steps in the framework, we consider the cost to perform the initial configuration optimization and simulation of the 16-hour workday. For a manufacturing line with 20 stations and 51 operations, the optimization algorithm (under one set of parameters) takes 0.8 s on average, and the simulation runs at around 400x real-time speed (it takes approximately 2.4 minutes to simulate 16 hours of the manufacturing process).

\subsection{Disturbance Scenarios}

We introduce two scenarios to evaluate the performance of the reconfiguration framework under different disturbances. In scenario 1, Worker2 in Fig. \ref{fig:line_disturbance_handle_addon} slows down and spends 50\% more time on all its operations. The simulation results are shown in \tableref{} \ref{tab:reconfig_performance}. Without reconfiguration, the station time will increase by around 20 s under this scenario and the throughput in 16 hours will decrease to 962. In scenario 2, worker2 spends 200\% more time on their operations, which will lead to a roughly 80 s increase in the station time. The throughput in 16 hours will decrease to 480.

The scenarios in the case study primarily illustrate the impact of worker delays, yet the framework is versatile enough to accommodate any disturbance that leads to shifts in operation times, such as worker absences, equipment malfunctions, and machine delays.

\subsection{Disturbance Handling through Line Reconfiguration}

\begin{figure}[t!]
	\centering
	\includegraphics[width=1.0\linewidth]{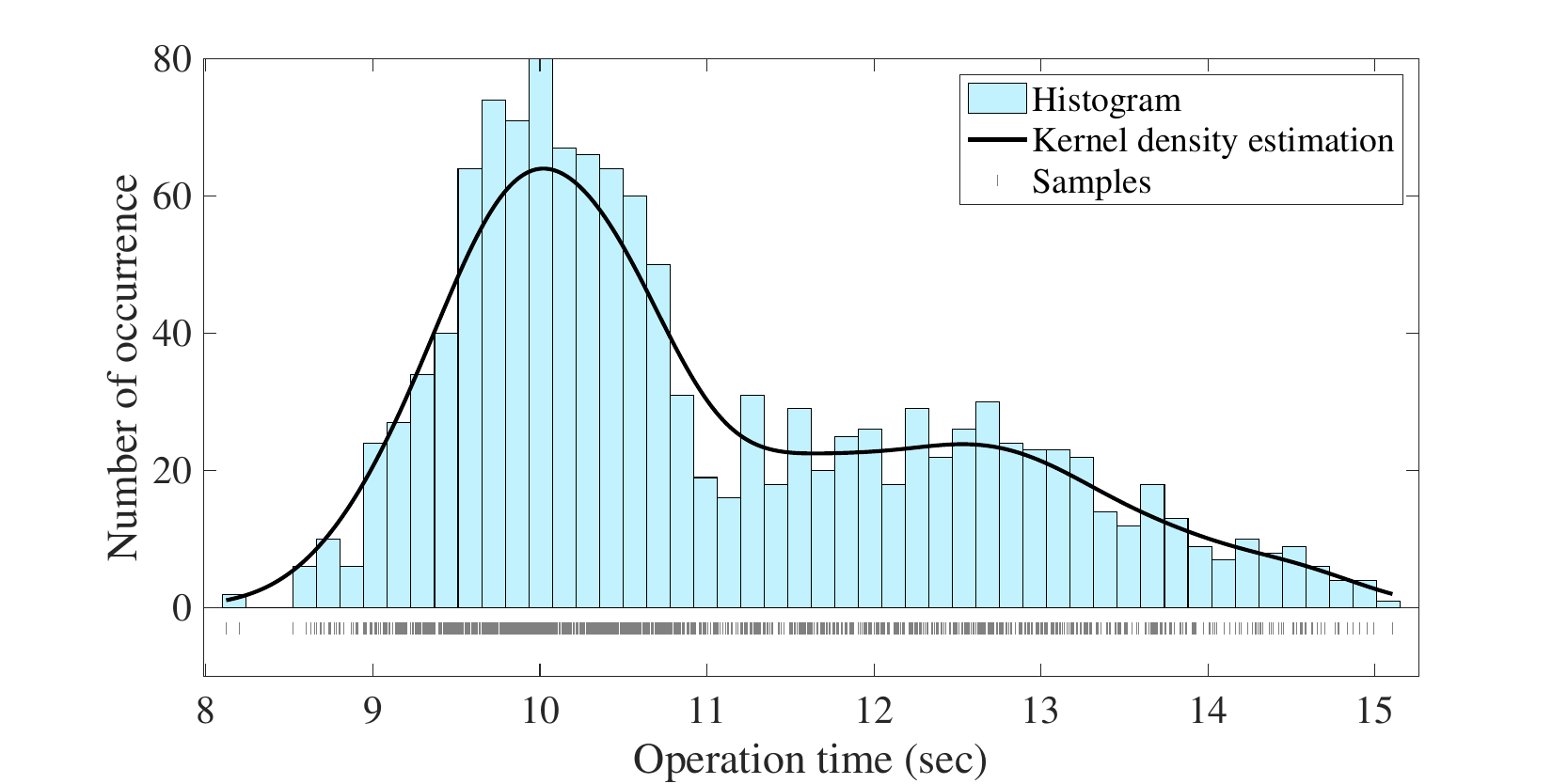}
	\caption{The time distribution for Worker2 to complete operation 25 when the worker is in a normal state. The histogram and density estimation are generated from data samples.}
	\label{fig:normal_worker_operation_time}
\end{figure}

The system process digital twin monitors the operation times of Worker2, and compares them with typical historical data (an example is shown in Fig. \ref{fig:normal_worker_operation_time}). When a disturbance occurs, the real-time operation time deviates from the typical time distribution. When the deviation is above a threshold, the digital twin triggers the reconfiguration framework in Fig. \ref{fig:system_flowchart}.
The ontology model is then queried to list the resources for all of the related operations and sends this information to the reconfiguration optimization.

During the reconfiguration optimization, we assume that adjacent agents of Worker2 include stations 5 (LargeRobot3), station 7 (WeldingMachine0), and station 8 (Worker3). Worker2 can either share its tasks with these neighbors or add additional agents that are not originally in the manufacturing line.

To illustrate the performance of our proposed line reconfiguration approach, we compare it with two alternative formulations. Our MILP approach allows sharing operations among multiple agents, so we first compare it with an MILP formulation that does not permit such sharing, which is the assumption used in MILP-based reconfiguration methods in~\cite{vahedi2022workforce, dou2020mixed, bortolini2021optimisation}, though the objectives and constraints differ. Additionally, we compare our approach with a state-of-the-art agent-based rescheduling method~\cite{bi2024dynamic}, which similarly allows sharing operations among agents during a disturbance and is well-suited to the scenarios considered in this case study. Notably, no existing line reconfiguration method can be directly applied to the mixed human-machine manufacturing scenarios addressed in this paper without violating underlying model assumptions; therefore, both baseline methods were modified to approximate the operational context of our study.

On average, our reconfiguration optimization takes 0.03~s to run. Compared to the configuration initialization, the time is shorter as the scope of the optimization is smaller. This scoped optimization partially modifies the line configuration under disturbances and improves the flexibility of the system.

For the larger disturbance (200\% increase in operation time), by changing the hyper-parameters, the optimizer exhibits the two major reconfiguration modes discussed in Sec. \ref{sec:line_config_optimizer}. The simulation results of the line configurations are listed in the last two rows in \tableref{} \ref{tab:reconfig_performance} for scenario 2. In the first mode (denoted as plan switch), agent usage is penalized more than cycle time, resulting in an optimization that selects sharing operation tasks with neighbors as the optimal reconfiguration. In the plan switch mode, the system is not able to maintain the original bottleneck time due to such a large disturbance and limited neighboring resources. In the second mode (denoted as configuration switch), maintaining the bottleneck time (cycle time) is heavily penalized, resulting in an optimization that includes the addition of one more robot to the manufacturing line to meet the original bottleneck time. However, this configuration introduces additional time and cost through the addition of a robot. 

For the smaller disturbance (50\% more time), the framework identified one optimal reconfiguration plan. Because the disturbance is relatively small, the optimizer is able to maintain the original bottleneck time without adding additional agents in the line (as demonstrated in the last row in \tableref{} \ref{tab:reconfig_performance} for scenario 1).

\begin{table}[t]
  % \centering
  \caption{Disturbance scenarios, reconfiguration modes, and corresponding performances*.}
  \resizebox{\columnwidth}{!}{%
    \begin{tabular}{c|c|ccc}
      \toprule
      \textbf{Scenario} & \textbf{Configuration} & \textbf{Agents} & \textbf{Bottleneck} & \textbf{Throughput} \\
      \midrule
      No disturbance & Original line         & 20    & 43.9 s  & 1295 \\
      \midrule
      \multirow{4}{*}{1: +50\% time} 
                     & Original line         & 20    & 59.3 s  & 962 \\
                     & MILP-based\(^\dagger\)            & 20    & 44.0 s  & 1289 \\
                     & Agent-based\(^\dagger\)           & 20    & 56.9 s  & 997  \\
                     & \textbf{Plan switch}  & \textbf{20} & \textbf{43.9 s}  & \textbf{1292} \\
      \midrule
      \multirow{5}{*}{2: +200\% time} 
                     & Original line         & 20    & 118.6 s & 480 \\
                     & MILP-based\(^\dagger\)            & 20    & 55.0 s  & 1032 \\
                     & Agent-based\(^\dagger\)           & 20    & 113.8 s & 500  \\
                     & \textbf{Plan switch}  & \textbf{20} & \textbf{48.8 s}  & \textbf{1161} \\
                     & \textbf{Config Switch} & \textbf{21} & \textbf{43.9 s}  & \textbf{1275} \\
      \bottomrule
    \end{tabular}%
  }
\(^\dagger\)Baseline methods.

*The table contains the number of agents used, average bottleneck time, and simulated throughput (in 16 hours) of the original line configuration and the configurations under different disturbances. The MILP-based and agent-based~\cite{bi2024dynamic} configurations are generated by the baseline methods we compared. The rows with bold text represent our proposed method.
\label{tab:reconfig_performance}%
\end{table}

The simulation generation program generates the setup for the modes (line configurations) in scenarios 1 and 2. The bottleneck times and the throughput values for the different conditions are listed in \tableref{} \ref{tab:reconfig_performance}.
For both scenarios 1 and 2, the two baseline methods handle disturbances through task redistribution (without adding new agents to the line). The MILP-based baseline does not allow sharing of any single operation, limiting reconfiguration flexibility, whereas the agent-based baseline prioritizes preserving the original production schedule over minimizing bottleneck time. As a result, both methods yield higher bottleneck times and lower throughput than our plan switch results, as shown in \tableref{} \ref{tab:reconfig_performance}. These results highlight the improved performance of our flexible line reconfiguration compared to the baselines.

Applying our configuration selector, when maintaining the bottleneck time is emphasized, the reconfiguration mode that includes an additional agent as the final plan for scenario 2 is selected. For scenario 1, since the optimizer returns only one optimal plan, that plan is selected. The modified configurations are illustrated in Fig. \ref{fig:line_disturbance_handle}b-c.

\begin{figure}[t!]
	\centering
	\includegraphics[width=1.0\linewidth]{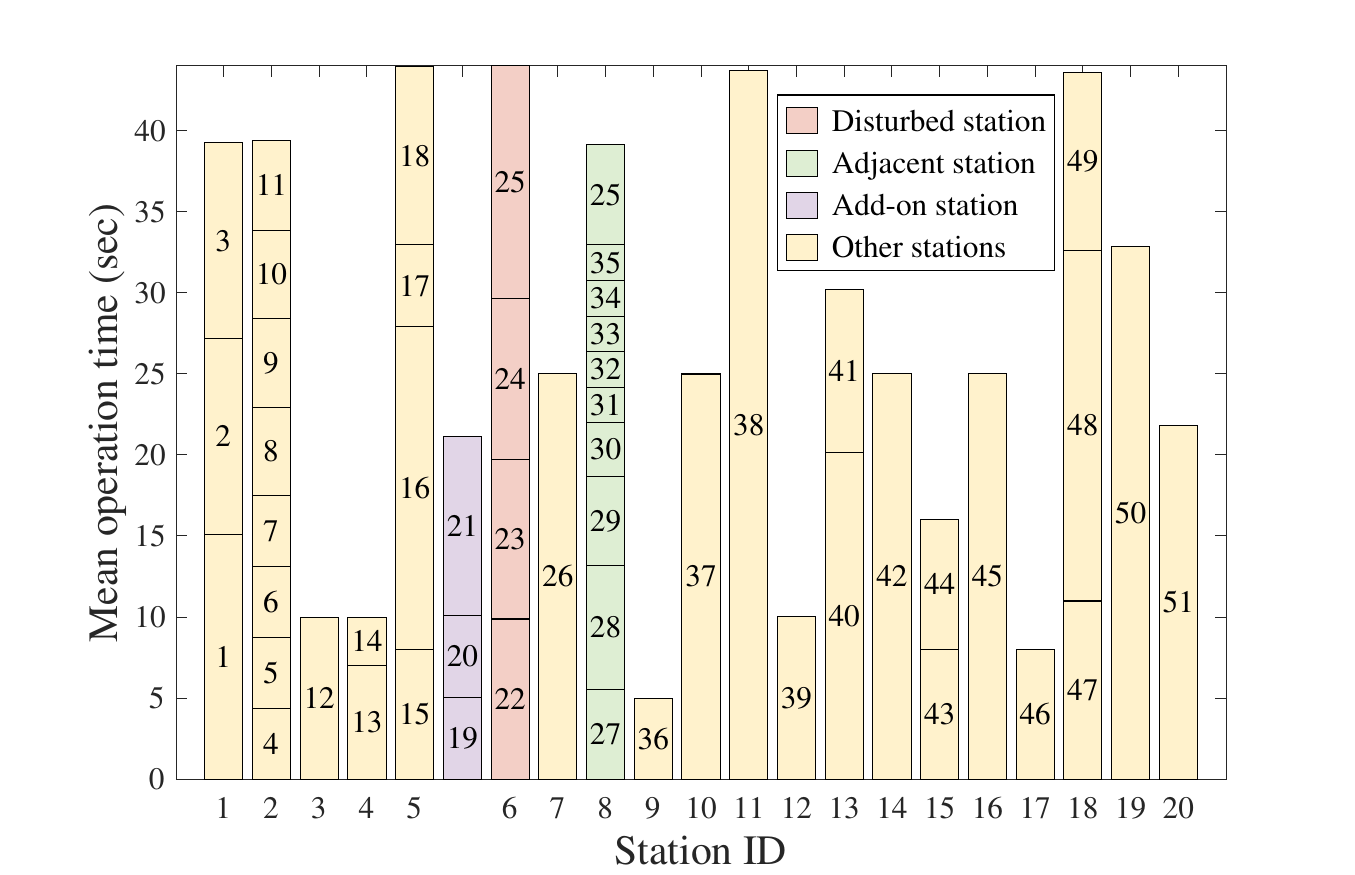}
	\caption{Configuration switch, disturbance scenario 2: the mean station and operation times according to the simulation. The corresponding line configuration is in Fig. \ref{fig:line_disturbance_handle_addon}. Each bar shows the total operation time of a station, while the smaller blocks within a bar indicate the operations within a station. In the blocks are the operation numbers. In this scenario, the worker at station 7 (red) delayed its operations. Therefore, 57\% of operation 25 is redistributed to the adjacent worker at station 8 (green). In addition, operations 19-21 are reassigned to a new station (purple) inserted between stations 5 and 6.}
	\label{fig:addon_station_operation_time}
\end{figure}

The selected reconfiguration plans (\tableref{} \ref{tab:reconfig_performance}, the last two rows in scenarios 1 and 2) successfully maintain the original bottleneck time of 43.9~s under both disturbance scenarios.
This indicates that, on average, one product is generated every 43.9~s, compared to 59.3~s and 118.6~s without reconfiguration in the two disturbance scenarios, respectively. A comparison of the time distributions for generating a single product, with and without reconfiguration, was conducted using a t-test. The resulting p-values in both scenarios were below 0.001, indicating statistically significant reductions in bottleneck time achieved through reconfiguration.
As a result, their throughput values in 16 hours increase to 1292 and 1275 products, respectively.
The implementation of a reconfiguration plan recovers the throughput, preventing the 26\% and 63\% drops that occur when disturbances arise but no reconfiguration plan is implemented (as shown in \tableref{} \ref{tab:reconfig_performance}).
Compared to the original scenario given in Fig. \ref{fig:initial_station_operation_time} and row 1 in \tableref{} \ref{tab:reconfig_performance}, the reconfiguration plans provide a feasible solution for handling the disruption to the manufacturing agent. 
When we compare the operation and station times of the original line (Fig. \ref{fig:initial_station_operation_time}), with the configuration switch (last row in \tableref{} \ref{tab:reconfig_performance}, shown in Fig. \ref{fig:addon_station_operation_time}), the reconfiguration plan exhibits one additional bottleneck station due to the disturbed agent.

According to the configuration shown in Fig. \ref{fig:line_disturbance_handle_addon}, worker3 performs both its own tasks at station 8 and operation 25 at station 6. In this scenario, the worker needs to move between two stations. This motivates our requirement that the task redistribution can only occur locally to adjacent agents. To ensure the feasibility of partial task assignments, the arrangement of adjacent stations should prioritize a reduced physical distance between stations. Furthermore, additional buffers should be added before operation 25. In this case, the worker would only need to transition between tasks when the buffer is full, and a larger buffer size reduces the number of transitions. In summary, allowing partial task assignment during reconfiguration increases flexibility but could introduce additional time costs due to the movement between stations. Future work will consider spatial constraints and any additional costs introduced by agent movements or redistributed tasks.

\subsection{Scalability of the Framework}

The proposed reconfiguration framework has demonstrated its applicability and efficiency in a case study featuring a manufacturing line with 51 operations and 8 agent types. This section discusses the scalability and potential challenges when applying the framework to larger manufacturing systems.

\subsubsection{Scalability of the Ontology Model, Configuration Optimizer, and Automatic Simulation}
The ontology model is designed to handle more complex relationships between agents and operations in larger-scale manufacturing lines. Similarly, the automatic simulation process scales well, as all simulation components are generated programmatically. The line configuration initialization currently takes an average of 0.8 s; although this may increase for larger lines, it can be computed offline and is not time-sensitive. The line reconfiguration optimization currently takes 0.03 s on average and has proven scalable for larger problems, as it relies on a linear program.

\subsubsection{System Process Digital Twin and Final Line Integration}

As manufacturing lines grow in length and complexity, synchronizing sensor data updates in the system process digital twin becomes more challenging. Moreover, modifying the manufacturing line through the line integrator (shown in Fig. \ref{fig:system_flowchart}) may take longer. A centralized approach to synchronize the information updates can be difficult to maintain, and delays in reconfiguration actions may increase. These issues can be mitigated by employing hierarchical or distributed strategies for line reconfiguration.

\subsubsection{Handling Multiple Concurrent Disturbances}
In larger, more complex manufacturing systems, multiple disturbances are more likely to occur simultaneously. Addressing these scenarios will require additional models to account for multi-point disturbances and ensure effective reconfiguration under concurrent disruptions.

\section{Conclusions and Future Work}\label{sec:conclusion}

This paper presents an integrated framework to initialize, adjust, and automatically verify manufacturing line configurations for disturbance handling in smart manufacturing. 
A system process digital twin and a capability-based ontology model are leveraged during the reconfiguration.
An innovative mixed-integer linear program-based disturbance handling configuration optimizer and a discrete-event simulation generation program are developed.
With these components, the framework identifies disturbances in an operating manufacturing line, stores/retrieves options and requirements for reconfiguration, optimizes throughput and agent usage, and rapidly evaluates system performances.

To demonstrate the efficacy of the framework, 
a battery manufacturing line-based case study with 51 operations and 8 agent types has been conducted. The framework initializes a line configuration to achieve the optimal bottleneck time of 43.9 s using 20 agents.
Then, we introduce two disturbances to the manufacturing line. 
Under the smaller disturbance, the reconfiguration framework is able to maintain the 43.9 s bottleneck time by redistributing operations from the disturbed agent to an adjacent agent with similar capabilities. Under larger disturbances, the framework chooses to add one more agent to the manufacturing line in addition to redistributing the operations to neighboring agents. Using these two strategies, the framework is able to determine a reconfiguration plan that maintains the original bottleneck time and throughput value. Compared to the use of the original line configuration plan under these two disturbances, the implementation of a reconfiguration plan prevents the 26\% and 63\% drops in the throughput, respectively.

In addition, the framework has been shown to be computationally efficient. The configuration optimization and reconfiguration take 0.8 s and 0.03 s on average to find an optimal solution. The simulation runs at around 400x real-time speed.

Future work will investigate additional considerations within the reconfiguration plans such as spatial constraints, experience levels, and retooling requirements that were not included in this work. Experimental validation on a manufacturing-like testbed will also be explored.

\bibliography{ms}

\begin{thebibliography}{10}
\providecommand{\url}[1]{#1}
\csname url@samestyle\endcsname
\providecommand{\newblock}{\relax}
\providecommand{\bibinfo}[2]{#2}
\providecommand{\BIBentrySTDinterwordspacing}{\spaceskip=0pt\relax}
\providecommand{\BIBentryALTinterwordstretchfactor}{4}
\providecommand{\BIBentryALTinterwordspacing}{\spaceskip=\fontdimen2\font plus
\BIBentryALTinterwordstretchfactor\fontdimen3\font minus
  \fontdimen4\font\relax}
\providecommand{\BIBforeignlanguage}[2]{{%
\expandafter\ifx\csname l@#1\endcsname\relax
\typeout{** WARNING: IEEEtran.bst: No hyphenation pattern has been}%
\typeout{** loaded for the language `#1'. Using the pattern for}%
\typeout{** the default language instead.}%
\else
\language=\csname l@#1\endcsname
\fi
#2}}
\providecommand{\BIBdecl}{\relax}
\BIBdecl

\bibitem{leitao2009agent}
P.~Leit{\~a}o, ``Agent-based distributed manufacturing control: A
  state-of-the-art survey,'' \emph{Engineering applications of artificial
  intelligence}, vol.~22, no.~7, pp. 979--991, 2009.

\bibitem{bi2023distributed}
M.~Bi, D.~M. Tilbury, S.~Shen, and K.~Barton, ``A distributed approach for
  agile supply chain decision-making based on network attributes,'' \emph{IEEE
  Transactions on Automation Science and Engineering}, vol.~21, no.~3, pp.
  2223--2236, 2023.

\bibitem{koren2010design}
Y.~Koren and M.~Shpitalni, ``Design of reconfigurable manufacturing systems,''
  \emph{Journal of manufacturing systems}, vol.~29, no.~4, pp. 130--141, 2010.

\bibitem{koren2018reconfigurable}
Y.~Koren, X.~Gu, and W.~Guo, ``Reconfigurable manufacturing systems:
  Principles, design, and future trends,'' \emph{Frontiers of Mechanical
  Engineering}, vol.~13, pp. 121--136, 2018.

\bibitem{yelles2021reconfigurable}
A.~R. Yelles-Chaouche, E.~Gurevsky, N.~Brahimi, and A.~Dolgui, ``Reconfigurable
  manufacturing systems from an optimisation perspective: a focused review of
  literature,'' \emph{International Journal of Production Research}, vol.~59,
  no.~21, pp. 6400--6418, 2021.

\bibitem{leng2022digital}
J.~Leng, Z.~Chen, W.~Sha, Z.~Lin, J.~Lin, and Q.~Liu, ``Digital twins-based
  flexible operating of open architecture production line for individualized
  manufacturing,'' \emph{Advanced Engineering Informatics}, vol.~53, p. 101676,
  2022.

\bibitem{bi2021dynamic}
M.~Bi, I.~Kovalenko, D.~M. Tilbury, and K.~Barton, ``Dynamic resource
  allocation using multi-agent control for manufacturing systems,''
  \emph{IFAC-PapersOnLine}, vol.~54, no.~20, pp. 488--494, 2021.

\bibitem{bi2024dynamic}
------, ``Dynamic distributed decision-making for resilient resource
  reallocation in disrupted manufacturing systems,'' \emph{International
  Journal of Production Research}, vol.~62, no.~5, pp. 1737--1757, 2024.

\bibitem{poudel2022integrated}
L.~Poudel, I.~Kovalenko, R.~Geng, M.~Takaharu, Y.~Nonaka, N.~Takahiro,
  U.~Shota, D.~M. Tilbury, and K.~Barton, ``An integrated framework for dynamic
  manufacturing planning to obtain new line configurations,'' in \emph{2022
  IEEE 18th International Conference on Automation Science and Engineering
  (CASE)}.\hskip 1em plus 0.5em minus 0.4em\relax IEEE, 2022, pp. 328--334.

\bibitem{kovalenko2022toward}
I.~Kovalenko, J.~Moyne, M.~Bi, E.~C. Balta, W.~Ma, Y.~Qamsane, X.~Zhu, Z.~M.
  Mao, D.~M. Tilbury, and K.~Barton, ``Toward an automated learning control
  architecture for cyber-physical manufacturing systems,'' \emph{IEEE Access},
  vol.~10, pp. 38\,755--38\,773, 2022.

\bibitem{yang2022intelligent}
S.~Yang and Z.~Xu, ``Intelligent scheduling and reconfiguration via deep
  reinforcement learning in smart manufacturing,'' \emph{International Journal
  of Production Research}, vol.~60, no.~16, pp. 4936--4953, 2022.

\bibitem{vahedi2022workforce}
B.~Vahedi-Nouri, R.~Tavakkoli-Moghaddam, Z.~Hanz{\'a}lek, and A.~Dolgui,
  ``Workforce planning and production scheduling in a reconfigurable
  manufacturing system facing the covid-19 pandemic,'' \emph{Journal of
  Manufacturing Systems}, vol.~63, pp. 563--574, 2022.

\bibitem{dou2020mixed}
J.~Dou, C.~Su, and X.~Zhao, ``Mixed integer programming models for concurrent
  configuration design and scheduling in a reconfigurable manufacturing
  system,'' \emph{Concurrent Engineering}, vol.~28, no.~1, pp. 32--46, 2020.

\bibitem{yelles2022minimizing}
A.~R. Yelles-Chaouche, E.~Gurevsky, N.~Brahimi, and A.~Dolgui, ``Minimizing
  task reassignments under balancing multi-product reconfigurable manufacturing
  lines,'' \emph{Computers \& Industrial Engineering}, vol. 173, p. 108660,
  2022.

\bibitem{nakano2021manufacturing}
T.~Nakano, K.~Daiki, H.~Chen, I.~Kovalenko, E.~Balta, Y.~Qamsane, and
  K.~Barton, ``Manufacturing line design configuration with optimized resource
  groups,'' in \emph{2021 IEEE 17th International Conference on Automation
  Science and Engineering (CASE)}.\hskip 1em plus 0.5em minus 0.4em\relax IEEE,
  2021, pp. 625--632.

\bibitem{bortolini2021optimisation}
M.~Bortolini, E.~Ferrari, F.~G. Galizia, and A.~Regattieri, ``An optimisation
  model for the dynamic management of cellular reconfigurable manufacturing
  systems under auxiliary module availability constraints,'' \emph{Journal of
  manufacturing systems}, vol.~58, pp. 442--451, 2021.

\bibitem{qamsane2019unified}
Y.~Qamsane, C.-Y. Chen, E.~C. Balta, B.-C. Kao, S.~Mohan, J.~Moyne, D.~Tilbury,
  and K.~Barton, ``A unified digital twin framework for real-time monitoring
  and evaluation of smart manufacturing systems,'' in \emph{2019 IEEE 15th
  international conference on automation science and engineering (CASE)}.\hskip
  1em plus 0.5em minus 0.4em\relax IEEE, 2019, pp. 1394--1401.

\bibitem{son2001automatic}
Y.~J. Son and R.~A. Wysk, ``Automatic simulation model generation for
  simulation-based, real-time shop floor control,'' \emph{Computers in
  Industry}, vol.~45, no.~3, pp. 291--308, 2001.

\bibitem{martinez2018automatic}
G.~S. Mart{\'\i}nez, S.~Sierla, T.~Karhela, and V.~Vyatkin, ``Automatic
  generation of a simulation-based digital twin of an industrial process
  plant,'' in \emph{IECON 2018-44th Annual Conference of the IEEE Industrial
  Electronics Society}.\hskip 1em plus 0.5em minus 0.4em\relax IEEE, 2018, pp.
  3084--3089.

\bibitem{delorme2024modelling}
X.~Delorme, G.~Fleury, P.~Lacomme, and D.~Lamy, ``Modelling and solving
  approaches for scheduling problems in reconfigurable manufacturing systems,''
  \emph{International Journal of Production Research}, vol.~62, no.~7, pp.
  2683--2704, 2024.

\bibitem{kombaya2022digital}
J.~Kombaya~Touckia, N.~Hamani, and L.~Kermad, ``Digital twin framework for
  reconfigurable manufacturing systems (rmss): design and simulation,''
  \emph{The International Journal of Advanced Manufacturing Technology}, vol.
  120, no.~7, pp. 5431--5450, 2022.

\bibitem{arnarson2023towards}
H.~Arnarson, H.~Yu, M.~M. Olavsbr{\aa}ten, B.~A. Bremdal, and B.~Solvang,
  ``Towards smart layout design for a reconfigurable manufacturing system,''
  \emph{Journal of Manufacturing Systems}, vol.~68, pp. 354--367, 2023.

\bibitem{caesar2023digital}
B.~Caesar, K.~Barton, D.~M. Tilbury, and A.~Fay, ``Digital twin framework for
  reconfiguration management: concept \& evaluation,'' \emph{IEEE Access},
  vol.~11, pp. 127\,364--127\,387, 2023.

\bibitem{leng2020digital}
J.~Leng, Q.~Liu, S.~Ye, J.~Jing, Y.~Wang, C.~Zhang, D.~Zhang, and X.~Chen,
  ``Digital twin-driven rapid reconfiguration of the automated manufacturing
  system via an open architecture model,'' \emph{Robotics and
  computer-integrated manufacturing}, vol.~63, p. 101895, 2020.

\bibitem{zhang2019reconfigurable}
C.~Zhang, W.~Xu, J.~Liu, Z.~Liu, Z.~Zhou, and D.~T. Pham, ``A reconfigurable
  modeling approach for digital twin-based manufacturing system,''
  \emph{Procedia Cirp}, vol.~83, pp. 118--125, 2019.

\bibitem{mo2023framework}
F.~Mo, H.~U. Rehman, F.~M. Monetti, J.~C. Chaplin, D.~Sanderson, A.~Popov,
  A.~Maffei, and S.~Ratchev, ``A framework for manufacturing system
  reconfiguration and optimisation utilising digital twins and modular
  artificial intelligence,'' \emph{Robotics and Computer-Integrated
  Manufacturing}, vol.~82, p. 102524, 2023.

\bibitem{pansare2023reconfigurable}
R.~Pansare, G.~Yadav, and M.~R. Nagare, ``Reconfigurable manufacturing system:
  a systematic review, meta-analysis and future research directions,''
  \emph{Journal of Engineering, Design and Technology}, vol.~21, no.~1, pp.
  228--265, 2023.

\bibitem{moyne2020requirements}
J.~Moyne, Y.~Qamsane, E.~C. Balta, I.~Kovalenko, J.~Faris, K.~Barton, and D.~M.
  Tilbury, ``A requirements driven digital twin framework: Specification and
  opportunities,'' \emph{Ieee Access}, vol.~8, pp. 107\,781--107\,801, 2020.

\bibitem{matsokis2010ontology}
A.~Matsokis and D.~Kiritsis, ``An ontology-based approach for product lifecycle
  management,'' \emph{Computers in industry}, vol.~61, no.~8, pp. 787--797,
  2010.

\bibitem{leitao2016industrial}
P.~Leit{\~a}o, A.~W. Colombo, and S.~Karnouskos, ``Industrial automation based
  on cyber-physical systems technologies: Prototype implementations and
  challenges,'' \emph{Computers in industry}, vol.~81, pp. 11--25, 2016.

\bibitem{jarvenpaa2019development}
E.~J{\"a}rvenp{\"a}{\"a}, N.~Siltala, O.~Hylli, and M.~Lanz, ``The development
  of an ontology for describing the capabilities of manufacturing resources,''
  \emph{Journal of Intelligent Manufacturing}, vol.~30, no.~2, pp. 959--978,
  2019.

\bibitem{kovalenko2019model}
I.~Kovalenko, D.~Tilbury, and K.~Barton, ``The model-based product agent: A
  control oriented architecture for intelligent products in multi-agent
  manufacturing systems,'' \emph{Control Engineering Practice}, vol.~86, pp.
  105--117, 2019.

\bibitem{qamsane2021methodology}
Y.~Qamsane, J.~Moyne, M.~Toothman, I.~Kovalenko, E.~C. Balta, J.~Faris, D.~M.
  Tilbury, and K.~Barton, ``A methodology to develop and implement digital twin
  solutions for manufacturing systems,'' \emph{IEEE Access}, vol.~9, pp.
  44\,247--44\,265, 2021.

\bibitem{toothman2023digital}
M.~Toothman, B.~Braun, S.~J. Bury, J.~Moyne, D.~M. Tilbury, Y.~Ye, and
  K.~Barton, ``A digital twin framework for prognostics and health
  management,'' \emph{Computers in Industry}, vol. 150, p. 103948, 2023.

\bibitem{wan2018ontology}
J.~Wan, B.~Yin, D.~Li, A.~Celesti, F.~Tao, and Q.~Hua, ``An ontology-based
  resource reconfiguration method for manufacturing cyber-physical systems,''
  \emph{IEEE/ASME Transactions on Mechatronics}, vol.~23, no.~6, pp.
  2537--2546, 2018.

\bibitem{musen2015protege}
M.~A. Musen, ``The prot{\'e}g{\'e} project: a look back and a look forward,''
  \emph{AI matters}, vol.~1, no.~4, pp. 4--12, 2015.

\bibitem{trappey2016review}
A.~J. Trappey, C.~V. Trappey, U.~H. Govindarajan, J.~J. Sun, and A.~C. Chuang,
  ``A review of technology standards and patent portfolios for enabling
  cyber-physical systems in advanced manufacturing,'' \emph{Ieee Access},
  vol.~4, pp. 7356--7382, 2016.

\bibitem{iso21838}
S.~International Organization~for Standardization~Geneva, \emph{Information
  technology — Top-level ontologies (TLO)}, International Organization for
  Standardization Geneva, Switzerland Std. ISO/IEC 21\,838, 2021.

\bibitem{iso2020automation}
------, \emph{Automation systems and integration—digital twin framework for
  manufacturing}, International Organization for Standardization Geneva,
  Switzerland Std. ISO/DIS 23\,247, 2020.

\end{thebibliography}
\bibliographystyle{IEEEtran}

\begin{IEEEbiography}[{\includegraphics
[width=1in,height=1.25in, trim=0 30 0 0,clip, keepaspectratio]{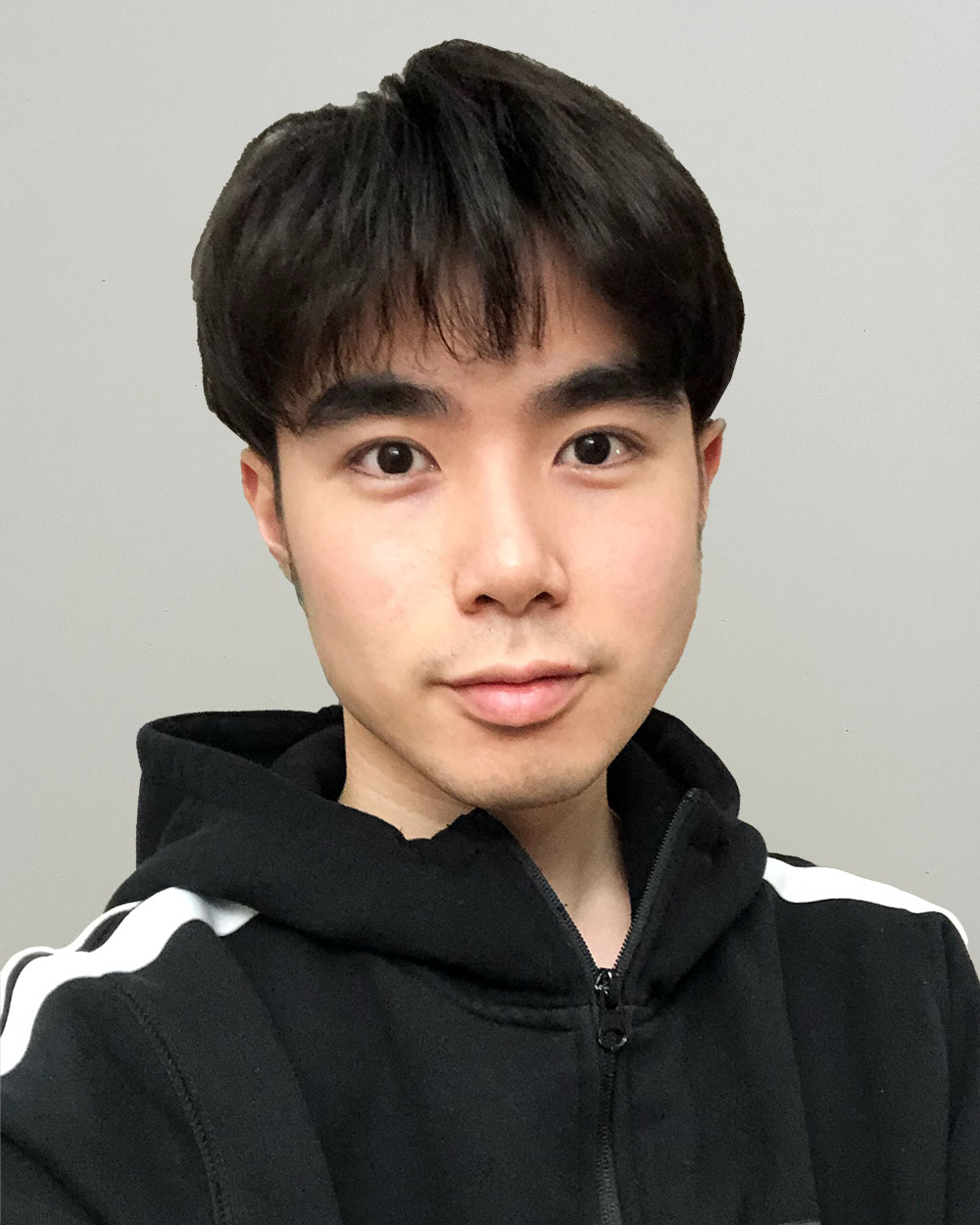}}] % <left> <lower> <right> <upper>
{Bo Fu} received the B.S. degree in vehicle engineering from Tongji University, China, in 2017, the M.S. degree in mechanical engineering from Carnegie Mellon University in 2019, and the Ph.D. degree in robotics from the University of Michigan in 2024. He is currently an Applied Scientist with Amazon Robotics. His research interests include robust and resilient decision-making for heterogeneous multi-agent systems under uncertainty using optimization and learning-based approaches. The applications include smart manufacturing systems, large-scale
multi-robot collaboration, human-robot systems, and smart buildings.
\end{IEEEbiography}

\begin{IEEEbiography}[{\includegraphics[width=1in,height=1.25in,clip,keepaspectratio]{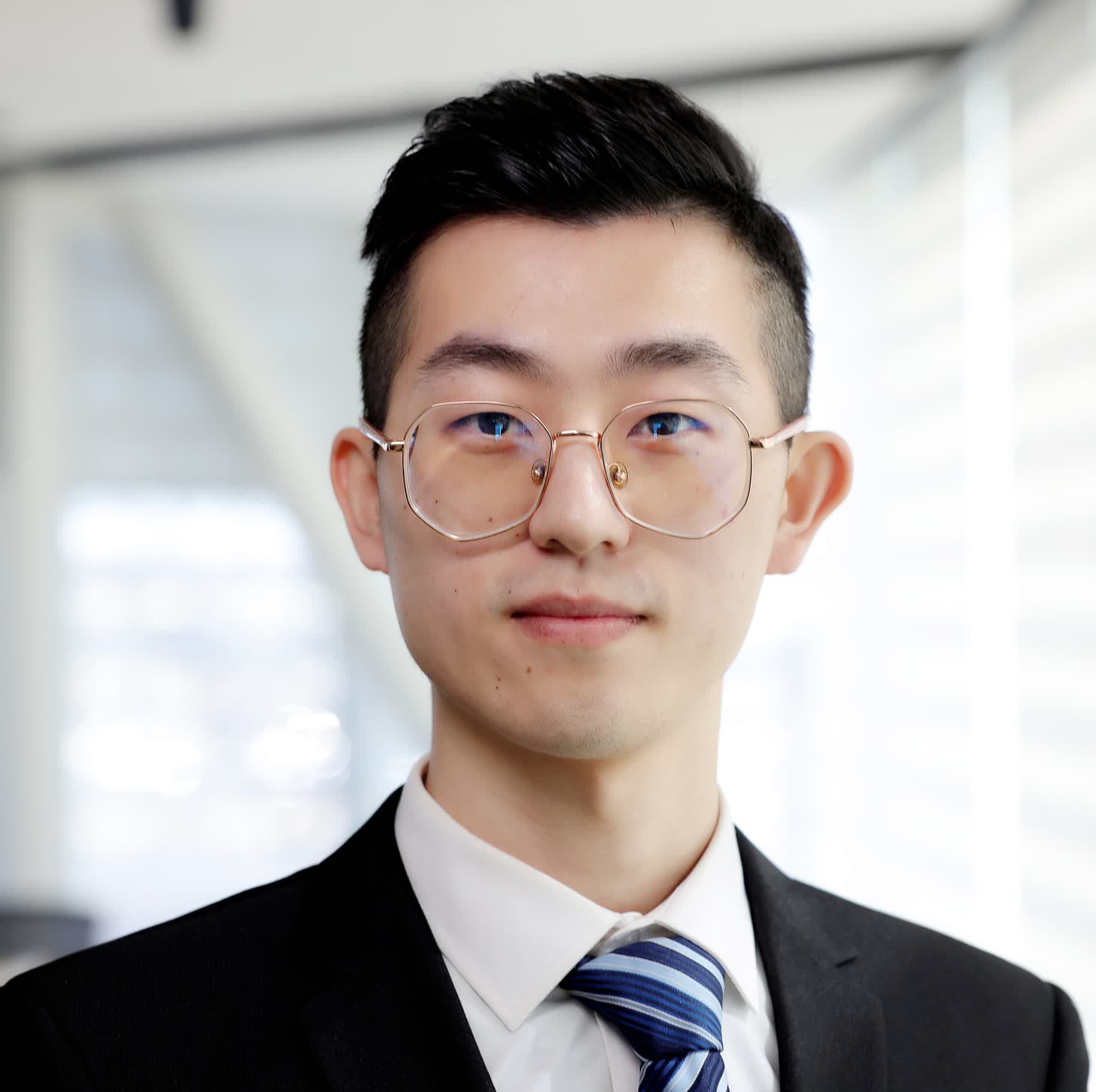}}]{Mingjie Bi} (Student Member, IEEE) received the B.S. degree in marine engineering from the Huazhong University of Science and Technology, China, in 2018, and the M.S. degree in mechanical engineering and the Ph.D. degree in robotics from the University of Michigan, Ann Arbor, MI, USA, in 2020 and 2023, respectively. He was a Smart Manufacturing Researcher with Hitachi America, Ltd. He is currently a Multi-Agent Research Scientist with Beijing Institute for General Artificial Intelligence. His research interests include multi-agent systems, swarm intelligence, smart industry, and robotics.
\end{IEEEbiography}

\begin{IEEEbiography}[{\includegraphics[width=1in,height=1.25in,clip,keepaspectratio]{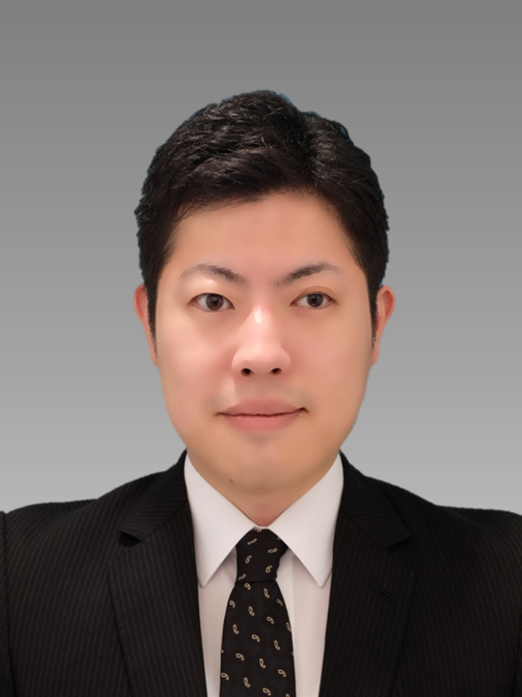}}]
{Shota Umeda} received the B.S. and M.S. degrees in mechanical engineering from Tokyo Institute of Technology, Tokyo, Japan, in 2014 and 2016, respectively. He joined the Yokohama Research Laboratory, Hitachi, Ltd., Yokohama, Japan, in 2016. He has been engaged in the application of statistical analysis and mechanical learning approaches to semiconductor plasma etchers. His research interests include flexible manufacturing systems and cyber-physical production systems (CPPS).
\end{IEEEbiography}

\begin{IEEEbiography}[{\includegraphics[width=1in,height=1.25in,clip,keepaspectratio]{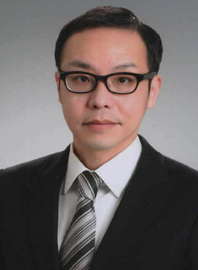}}]
{Takahiro Nakano} received the B.S. degree,  M.S. degree, and Ph.D. degree from Kyushu Institute of Technology in 2001, 2003, and 2006, respectively. In 2006, he joined Hitachi Ltd. Research and Development Group and is currently the manager of a research unit on smart manufacturing and robot production systems. From 2012 to 2013, he served as a visiting researcher at the Hungarian Academy of Sciences. He received the Precision Engineering Society's Technical Award in 2017. His research interests include digital engineering, production line engineering, production scheduling, computer-aided process planning (CAPP), and cyber-physical production systems (CPPS).
\end{IEEEbiography}

\begin{IEEEbiography}[{\includegraphics[width=1in,height=1.25in,clip,keepaspectratio]{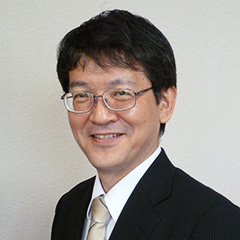}}]
{Youichi Nonaka} received the B.S. degree from University of Tsukuba, the M.S. degree from Tohoku University, and the Ph.D. degree from Tokyo Institute of Technology. He joined the Production Engineering Research Laboratory of Hitachi Ltd in 1992, working in R\&D for industrial robot application systems, digital engineering technology, and production control technology. He was a visiting researcher at the Massachusetts Institute of Technology in 2001. He has served as an international expert and Japan representative in convening ISO / IEC international standardization activities about Smart Manufacturing since 2014. Also, He has been a Collaborative Professor at the Graduate School of Informatics, Kyoto University, since 2018. He assumed his current position as Corporate Chief Researcher of the Research and Development Group of Hitachi Ltd in 2023.
\end{IEEEbiography}

\begin{IEEEbiography}[{\includegraphics[width=1in,height=1.25in,clip,keepaspectratio]{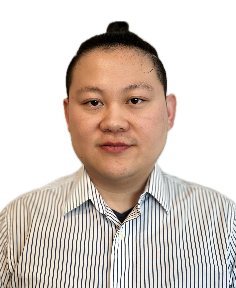}}]
{Quan “Jason” Zhou} received the B.S. degree from Chongqing University in 2012, and Ph.D. degree from Michigan State University in 2019 both in Materials Science and Engineering. After graduation, he worked at the Advanced Photon Source at the Argonne National Lab as a post-doctoral researcher with Purdue University and then the University of Utah. Jason joined the R\&D division of Hitachi America Ltd. as a Smart Manufacturing Researcher in 2020. His research interests include smart manufacturing, flexible manufacturing systems, industrial human digital twins, and applied ergonomics.
\end{IEEEbiography}

\begin{IEEEbiography}[{\includegraphics[width=1in,height=1.25in,clip,keepaspectratio]{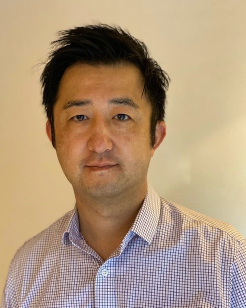}}]
{Takaharu Matsui} received the B.S. degree and the M.S. degree in social engineering from Tokyo Institute of Technology in 2000 and 2002, respectively. After graduation, he joined the Research and Development Group in Hitachi, Ltd., Japan, in 2002 as a researcher. He has 20+ years of research experience in manufacturing systems and production engineering. He also has various experiences in leading co-creation projects with external customers and internal Hitachi Group factories in various types of products and manufacturing. He is a member of the Japan Society of Mechanical Engineers. He holds 10+ patents through his research activities. He is currently a Senior Director of the Research and Development Division in Hitachi America, Ltd., US, leading global research and new business development projects for industrial digital solutions and services. His research interests include digital engineering, robotics automation, human digitization, industrial AI and optimization, human-machine interaction, cyber-physical systems, and cognitive systems in manufacturing and warehousing.
\end{IEEEbiography}

\begin{IEEEbiography}[{\includegraphics[width=1in,height=1.25in,clip,keepaspectratio]{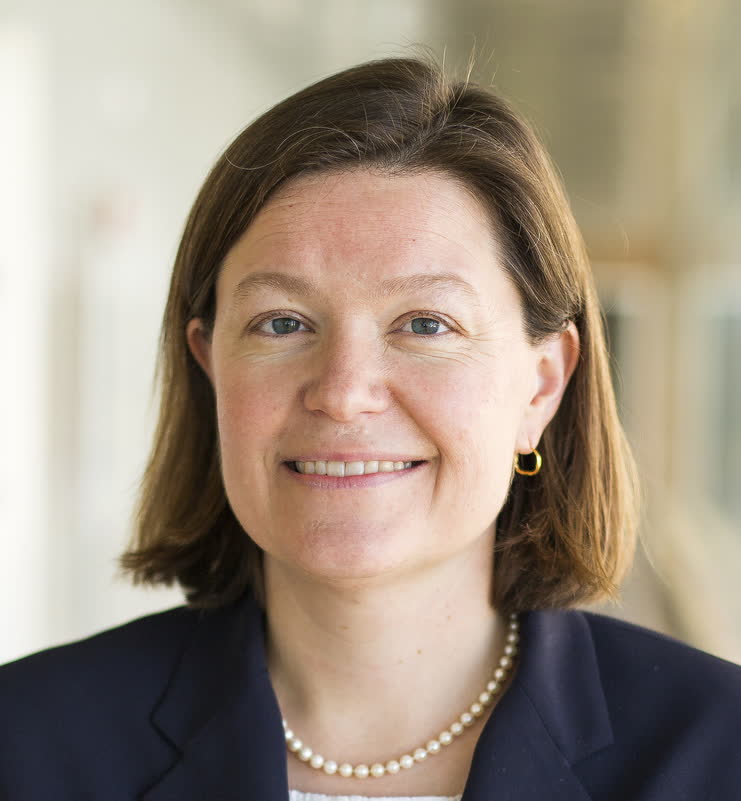}}]{Dawn M. Tilbury}
(Fellow, IEEE) is the inaugural Ronald D. and Regina C. McNeil Department Chair of Robotics at the University of Michigan, and the Herrick Professor of Engineering. She received the B.S. degree in Electrical Engineering from the University of Minnesota, and the M.S. and Ph.D. degrees in Electrical Engineering and Computer Sciences from the University of California, Berkeley.  Her research interests lie broadly in the area of control systems, including applications to robotics and manufacturing systems.  From 2017 to 2021, she was the Assistant Director for Engineering at the National Science Foundation, where she oversaw a federal budget of nearly \$1 billion annually, while maintaining her position at the University of Michigan. She has published more than 200 articles in refereed journals and conference proceedings.  She is a Fellow of IEEE, a Fellow of ASME, and a Life Member of SWE.
\end{IEEEbiography}

\vspace{-10cm}

\begin{IEEEbiography}[{\includegraphics
[width=1in,height=1.25in, trim=0 0 0 0,clip,
keepaspectratio]{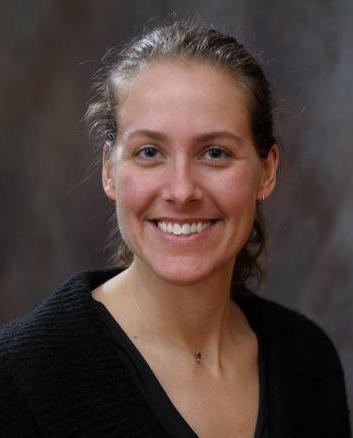}}] % <left> <lower> <right> <upper>
{Kira Barton} (Senior Member, IEEE) received her B.S. degree in Mechanical Engineering from the University of Colorado in 2001, and her M.S. and Ph.D. in Mechanical Engineering from the University of Illinois at Urbana-Champaign in 2006 and 2010. She joined the Mechanical Engineering Department at the University of Michigan, Ann Arbor in 2011. She is currently a Professor in the Robotics Department and Mechanical Engineering Department. She is also serving as the Associate Director for the Automotive Research Center, a University-based U.S. Army Center of Excellence for modeling and simulation of military and civilian ground systems. Prof. Barton’s research specializes in advancements in modeling, sensing, and control for applications in smart manufacturing and robotics. She is the recipient of an NSF CAREER Award in 2014, 2015 SME Outstanding Young Manufacturing Engineer Award, the 2015 University of Illinois, Department of Mechanical Science and Engineering Outstanding Young Alumni Award, the 2016 University of Michigan, Department of Mechanical Engineering Department Achievement Award, and the 2017 ASME Dynamic Systems and Control Young Investigator Award. She was named 1 of 25 leaders transforming manufacturing by SME in 2022, and was selected as one of the 2022 winners of the Manufacturing Leadership Award from the Manufacturing Leadership Council.
\end{IEEEbiography}

\end{document}